\documentclass [twocolumn]{aastex63}
\usepackage{savesym}
\usepackage{textcomp}
\usepackage{graphicx}
\usepackage{float}
\usepackage{bm}
\usepackage{amsmath}
\usepackage{mathtools}
\usepackage{multirow}
\usepackage{enumitem}
\usepackage{lineno}

\published{September 16, 2022}
\shorttitle{FFF Solver in Poloidal-Toroidal Representation}
\shortauthors{Yi et al.}

\begin{document}

\title{Reconstruction of Coronal Magnetic Fields Using a Poloidal-Toroidal Representation} 

\correspondingauthor{G.~S. Choe}
\email{gchoe@khu.ac.kr}

\author{Sibaek Yi}
\affiliation{School of Space Research, Kyung Hee University, Yongin 17104, Korea}

\author{G. S. Choe}
\affiliation{School of Space Research, Kyung Hee University, Yongin 17104, Korea}
\affiliation{Department of Astronomy and Space Science, Kyung Hee University, Yongin 17104, Korea}

\author{Kyung-Suk Cho}
\affiliation{Korea Astronomy and Space Science Institute, Daejeon  34055, Korea}

\author{Sami K. Solanki}
\affiliation{Max-Planck-Institut f\"{u}r Sonnensystemforschung, D-37077 G\"{o}ttingen, Germany}

\author{J\"{o}rg B\"{u}chner}
\affiliation{Zentrum f\"{u}r Astronomie und Astrophysik, Technische Universit\"{a}t Berlin, D-10623 Berlin, Germany}
\affiliation{Max-Planck-Institut f\"{u}r Sonnensystemforschung, D-37077 G\"{o}ttingen, Germany}

%
%
%
%
%
%
%



\begin{abstract}

A new method for reconstruction of coronal magnetic fields as force-free fields (FFFs) is presented. Our method employs poloidal and toroidal functions to describe divergence-free magnetic fields. This magnetic field representation naturally enables us to implement the boundary conditions at the photospheric boundary, i.e., the normal magnetic field and the normal current density there, in a straightforward manner.  At the upper boundary of the corona, a source-surface condition can be employed, which accommodates magnetic flux imbalance at the bottom boundary. Although our iteration algorithm is inspired by extant variational methods, it is non-variational and requires far fewer iteration steps than most of them. 
The computational code based on our new method is tested against the analytical FFF solutions by Titov \& D\'{e}moulin (1999). It is found to excel in reproducing a tightly wound 
flux rope, a bald patch and quasi-separatrix layers with a hyperbolic flux tube.

\end{abstract}

\keywords{Solar magnetic fields --- Solar corona --- Computational methods}


\section{Introduction} \label{sec:intro}

To understand physical processes in the solar corona and their connection with dynamics of the heliosphere, information about the coronal magnetic field is indispensable. 
Measurements of coronal magnetic fields have been made by use of radio observations \citep{Alissandrakis95, White91, Brosius97, Lee98, White05}, 
by spectropolarimetry of near infrared lines 
\citep{Lin04, Tomczyk08, Judge13, Plowman14, Dima20} or by utilizing a magnetic-induced transition line in Fe~X and other Fe~X and Fe~XI lines in EUV 
\citep{Landi20}. 
These techniques as of today, however, can only provide a coarse 2D map of the line-of-sight magnetic field, the magnitude of the magnetic field or a pointwise vector field. 
Coronal magnetic field strengths have also been indirectly estimated from observations of coronal loop oscillations \citep{Nakariakov01, VanDoorsselaere08} and of coronal mass ejections (CMEs) \citep{Jang09, Gopalswamy12}. These indirect measurements are too locally confined to give us a geometrical picture of the coronal magnetic field.  
On the other hand, the magnetic field vectors in the photosphere have been measured by spectropolarimetry with a sufficiently high resolution and accuracy
to produce a vector magnetogram, which provides a 2D cut near the solar surface through the 3D vector field \citep{Beckers71, Harvey85, Solanki93, Lites00}. 

With photospheric vector magnetograms available, there have been efforts to reconstruct coronal magnetic fields. Although the real corona is fully dynamic, the magnetic field that can be generated by one vector magnetogram without knowing its prehistory is a static field. A  magnetohydrostatic field under gravity requires the information of pressure 
(and temperature if not isothermal) at the photospheric level in addition to the vector magnetogram \citep{Grad58, Wiegelmann07, Zhu18}, 
but such information is not readily available yet. 
Since the plasma $\beta$, the ratio of plasma pressure and magnetic pressure, in the corona, 
especially in active regions, is much less than unity (\citealt{Iwai14}, cf. \citealt{Rodriguez19}), 
the approximation of coronal magnetic field to a force-free field (FFF) has been prevalent. 
An FFF in a domain $V$ is a vector field $\bm {B}$ such that 
\begin{align}
{\bm J} \times {\bm B} & = 0  
\ \Rightarrow \  
{\bm J} = \nabla \times {\bm B}  = \alpha ({\bm r}) {\bm B}\, , \label{eq:fff} \\
\nabla \cdot {\bm B}  & = 0\, . \label{eq:gauss}
\end{align}
As in equation~(\ref{eq:fff}), we will omit constant coefficients in Maxwell's equations by a proper normalization
throughout this paper. 
From equations (\ref{eq:fff}) and (\ref{eq:gauss}), we have 
\begin{equation} \label{eq:alpha}
 {\bm B} \cdot \nabla {\alpha} = 0 \, 
\end{equation}
i.e., $\alpha$ is constant along each field line. 
If the scalar field $\alpha$ is constant everywhere in $V$, equations (\ref{eq:fff}) and (\ref{eq:gauss}) form a linear vector Helmholtz equation
\begin{equation} \label{eq:Helmholtz}
 \nabla^2 {\bm B} + \alpha^2 {\bm B}=0  \, , 
\end{equation}
and its solution, an FFF with a constant $\alpha$, is a linear force-free field \citep{Aly92}. 
If $\alpha$ at the boundary $\partial V$ is non-constant, the scalar field $\alpha ({\bm r})$ for ${\bm r} \in V$ is an unknown, 
and such an FFF is a nonlinear force-free field (NLFFF). 
Since the scalar $\alpha$ in the photosphere is far from constant, reconstruction of a coronal magnetic field amounts to seeking an NLFFF. 

The earliest attempts of constructing an NLFFF used an algorithm, in which the three components of 
the vector field are computed from the photospheric boundary successively upward plane by plane \citep{Nakagawa74, Wu85,  Wu90} 
as if a time-dependent hyperbolic partial differential equation (e.g., an advection equation) were solved  marching along the time coordinate, 
whose role is taken by the vertical coordinate $z$ in the FFF solver.   
As already pointed out by \citet{Grad58}, a magnetohydrostatic (MHS) equation has imaginary characteristics like an elliptic equation (e.g., Poisson equation) as well as real characteristics like a hyperbolic equation.
Thus, solving a force-free equation, which is semi-elliptic, as a Cauchy problem is ill-posed so that
the successive integration  
algorithm cannot avoid exponentially growing errors as going up to a higher altitude \citep{Aly89, Amari97, McClymont97}. 
For this reason, the algorithm is now seldom used, but it has left a legacy of the much-used term ``extrapolation.'' 
In this paper, we will use the term ``reconstruction'' instead of the somewhat misleading ``extrapolation''  to refer to solving for a coronal magnetic field with certain boundary conditions. 

Among a variety of coronal NLFFF solvers that have been proposed so far \citep[for a review see][]{Aly07, Wiegelmann21}, 
there are two major groups of methods that are in practical these days. 
They are Grad-Rubin methods (current-field iteration methods) and variational methods. The former were originally proposed by \citet{Grad58} and have been applied 
in diverse formulations and algorithms \citep{Sakurai81, Amari99, Amari06, Wheatland06}. 
The Grad-Rubin methods in common employ an iteration procedure, in which the domain is first loaded with a field-aligned
electric current ${\bm J}^{m+1}=\alpha^{m} {\bm B}^{m}$
satisfying equation (\ref{eq:alpha}) for ${\bm B}^{m}$ at the iteration step $m$ and $\alpha = {\bm J} \cdot {\hat{\bm n}} / {\bm B} \cdot {\hat{\bm n}}$ 
specified at the boundary points (footpoints)
having one sign of the normal field component ${\bm B} \cdot {\hat{\bm n}}$ in $\partial V$, 
and then ${\bm B}^{m+1}$ is updated by solving the equation $ \nabla \times {\bm B}^{m+1} = {\bm J}^{m+1}$. 
Although the Grad-Rubin problem, which is to 
solve equations~(\ref{eq:fff})--(\ref{eq:gauss}) with $B_n$ given at every point of $\partial V$ and $\alpha$ only in the part of $\partial V$ with one sign of $B_n$, is known to
be well-posed for $|\alpha| <\alpha_m < \infty$  \citep{Bineau72, Boulmezaoud00}, it has not yet been rigorously proved whether the Grad-Rubin iteration procedure always converges to a solution or not.  
However, numerical codes based on the Grad-Rubin method have successfully been applied to real solar problems 
demonstrating its usefulness \citep{Bleybel02, Regnier04, Regnier07, Petrie11, Mastrano20}.  

In the variational methods, we solve for the magnetic field that extremizes (actually minimizes) a certain functional, 
which is usually a volume integral involving magnetic field, with certain boundary conditions and some additional constraints if any.  
For example, if a certain field line connectivity is imposed in $V$ and
the conjugate footpoints of each field line are fixed in $\partial V$,  and  
if the magnetic field in $V$ is varied under the ideal magnetohydrodynamic (MHD) condition without footpoint motions in $\partial V$
so that the first two constraints may be maintained, the magnetic field that minimizes the functional 
$\displaystyle W = {1 \over 2} \int_V {{\bm B}\cdot {\bm B}} \, dV$, 
the total magnetic energy in the domain $V$, is a force-free field \citep{Grad58, Chodura81}. 
A sufficient condition for $\delta W \le 0$ is that the fictitious plasma velocity is proportional to the Lorentz force, i.e., 
${\bm v} \propto {\bm J} \times {\bm B} $. 
This physically implies that the force-free state can be approached by removing the kinetic energy, into which the excessive potential energy is converted,  
from the system possibly by a hypothetical friction and/or viscous diffusion. 

However, the problem of reconstructing coronal magnetic field is quite different from that treated by \citet{Chodura81}. 
Since we do not know the field connectivity beforehand, we need to lift the ideal MHD condition and the constraint of field connectivity intentionally.
Instead we have to impose the normal current density $J_n$ or 
three components of $\bm B$ at $\partial V$. Then, an energy-decreasing evolution  
tends to deplete magnetic helicity through its dissipation within the system and 
transport through the boundary \citep{Berger84}. 
Thus, maintaining $J_n$ or ${\bm B}$ at $\partial V$ requires winding up field line footpoints there. 
Magnetofrictional methods for coronal FFF reconstruction either alternate stressing and relaxing steps explicitly \citep{Mikic94, Roumeliotis96, Jiao97} or are inherently 
equipped with 
a rather automatic re-stressing mechanism  \citep{Valori05, Valori10, Inoue14, Guo16, Jiang16}. 
Since the magnetofrictional codes are more or less modified forms of MHD solvers, 
they can employ free boundary conditions, also called open boundary conditions, at the outer boundary 
(lateral and top boundaries for a box-shaped domain) \citep[e.g.,][]{Valori10}. This flexibility in the outer boundary conditions helps the system to evolve
toward a force-free state in most cases, but it is uncertain what mathematical problem the resulting force-free state is the solution of.  
 
Another group of most widely used variational methods are the so-called optimization methods, in which the functional 
\begin{equation} \label{eq:Lopti}
L = \int_V \left[ {{ \left| {\bm J} \times {\bm B} \right|^2 } \over {B^2}} + \left( \nabla \cdot {\bm B} \right)^2 \right] dV
\end{equation}
is to be minimized \citep{Wheatland00}. 
The simplest choice of the boundary condition well-posing this variational problem is 
${\partial {\bm B}} / {\partial t}=0$, which can straightforwardly be implemented. 
There has been a concern that a boundary condition giving all three components of magnetic field results in an overspecified problem in contrast to 
the Grad-Rubin formulation \citep{Grad58, Boulmezaoud00}. However, available observational data of photospheric magnetic field and the boundary 
conditions employed at the outer boundary (lateral and top boundaries for a box-shaped domain) 
can hardly satisfy the compatibility relations of force-free fields \citep{Aly89, Wiegelmann21}. 
The mathematically clear formulation of the optimization methods tells that specifying all three components of magnetic field at the entire boundary works
for minimizing the functional $L$ in equation (\ref{eq:Lopti}) to a value, which is not necessarily zero, even if the compatibility conditions are 
not met. This robustness is a great merit of the optimization method as is its well-posedness as a variational problem. 
It is, however, a shortcoming of the method that the outcome strongly depends on what $\bm B$ is given at the outer boundary \citep{Wiegelmann06b}. 
It should be also noted that in the optimization method, the nonzero $\nabla \cdot {\bm B}$ term 
arises not merely from numerical discretization errors, but rather from the non-divergence-free 
form of ${\partial {\bm B}} / {\partial t}$ required for reducing the functional $L$ given above\footnote{ 
Also in magnetofrictional methods, nonzero $\nabla \cdot {\bm B}$ terms are often 
intentionally included in the induction 
equation to create parallel electric field to induce magnetic reconnection 
and in the momentum equation to remove magnetic charge out of the system \citep{Valori10, Inoue14}.
}. 
As a remedy for this concern, it has been suggested to impose $\nabla \cdot {\bm B}=0$
as a constraint using a Lagrange multiplier in the variational procedure \citep{Nasiri19}. 
It has also been tried to purge the final solution of nonzero  $\nabla \cdot {\bm B}$ by a postprocessing 
\citep{Rudenko20}, 
which, however, entails an unignorable change in either the normal or tangential component of magnetic field 
at the bottom boundary. 

Besides the two classes of methods above, the boundary integral method, also called boundary element method, 
has also been applied in the solar context \citep{He08, He20, Guo19}. This method is quite similar to the Green function method 
for partial differential equations, and a surface integral involving a reference function over the photospheric boundary needs to be evaluated 
to obtain $\bm B$ at each coronal point at every iteration step \citep{Yan00, Yan06}. 
It is a merit of the boundary integral method that a finite computational domain does not need to be set up, 
nor any artificial boundary conditions at the outer boundary, because it assumes a semi-infinite domain and a finite total magnetic energy in it. 

All the aforementioned methods solve an FFF problem with $B_n$ and $J_n$ or $\bm B$ as the bottom boundary condition. The
problems so posed aim to reconstruct 
coronal magnetic fields with vector magnetograms at hand. Our new method presented here also tackles such problems. 
However, FFF problems may be posed in different ways depending on one's interest. 
If it is necessary or desirable to impose a certain field connectivity, one can use a magnetofrictional method using Euler potentials \citep[e.g.,][]{Choe02} 
or the fluxon method using thin, piecewise linear flux tubes called fluxons for magnetic field description \citep{Deforest07}.  
In another problem setting, an FFF solution was sought with a flux rope initially placed at a desired location \citep{vanBallegooijen04, vanBallegooijen07}.  
Here, only $B_n$ is imposed at the bottom boundary and $J_n$ comes 
out of the solution.

With such a variety of methods for coronal FFF reconstruction available today, we still want to present a new method, which has
the following properties. 
\begin{enumerate}[label=(\arabic*)]
\item The magnetic field is described in a way ensuring divergence-freeness.  
\item The boundary conditions at the bottom boundary, $B_z (z=0)$, and $J_z (z=0)$, are straightforwardly implemented once and for all. 
\item The lateral and top boundary conditions can accommodate magnetic flux imbalance at the photospheric boundary. 
\item Fewer iteration steps are required for convergence than in most other methods. 
\item The numerical code is robust and equally operative for simple and complex field geometries. 
\end{enumerate}
Let us make some remarks on the above items. The most fundamental way of describing a divergence-free (solenoidal) vector field is using a vector potential such 
that ${\bm B}=\nabla \times {\bm A}$. Imposing $B_z$ at the boundary $z=0$ would be readily done by fixing appropriate $A_x$ and $A_y$ at the boundary once and for all. 
However, $J_z$ or $B_x$ and $B_y$ cannot be determined by the values of $\bm A$ at the boundary only. Therefore, we have to adjust some components of $\bm A$ at $z=0$
at every iteration step in keeping with the evolution inside the domain \citep[e.g.,][]{Roumeliotis96}. Thus, items (1) and (2) cannot be realized together 
when vector potentials being used.
In this paper, we describe magnetic field with two scalar functions $\Phi$ and $A_z$, named poloidal and toroidal functions, which will be explained in detail in Section~\ref{sec:pt}. 
Such a field description has never been tried in numerical computation of solar magnetic fields whether static (FFF included) or dynamic. 
In our formulation, the divergence-freeness of $\bm B$ is guaranteed, and $B_z$ is solely represented by $\Phi$, and $J_z$ by $A_z$. 
Thus, setting boundary conditions at $z=0$ is done once and for all. 

With regard to item (3), we note that the total positive and negative fluxes are 
generally not equal 
in magnitude in the magnetogram of any active region. The imbalanced fluxes are often coerced into achieving a balance by preprocessing \citep{Wiegelmann06a}, 
or are accommodated in FFF computations 
by assuming certain symmetries across the lateral boundaries \citep{Seehafer78, Otto07}. 
In our model, we set up the lateral boundaries as rigid conducting walls so that no magnetic flux may 
escape the domain through the boundaries and field lines tangential to them may slide freely during iterations. 
As for the top boundary, we employ a source surface 
condition, in which the magnetic field should have only the normal component ($B_z$), but no tangential components ($B_x=B_y=0$). Thus, the unpaired extra magnetic 
flux at the bottom boundary is connected to the top boundary so that the condition 
$\displaystyle \oint_S {\bm B}\cdot {\hat{\bm n} }\, dS=0$ 
should be met. The source surface boundary condition at the outer corona was first suggested by \citet{Altschuler69} and has since
been widely applied to potential field models. A theory of NLFFFs with a source surface boundary condition was put forward by \citet{Aly93}, 
but FFF reconstruction with $B_n$ and $J_n$ together or ${\bm B}$ as the bottom boundary condition and with the source surface condition at the top boundary 
has not been attempted before the present paper\footnote{
In \citet{vanBallegooijen00}, \citet{vanBallegooijen04} and \citet{vanBallegooijen07}, a source surface condition was employed at the top boundary, 
but it was not intended to solve an FFF problem, in which both $B_n$ and $J_n$ (or~${\bm B}$) are imposed as the bottom boundary conditions. 
}.

New NLFFF solvers must be tested against known analytical FFFs before being applied to solar vector magnetograms. 
Currently two analytical NLFFF solutions are widely used as reference fields. The FFF models by \citet{Low90} (hereafter LL) are exact analytical solutions, 
which involve modestly sheared magnetic fields without flux ropes. 
The models by \citet{Titov99} (hereafter TD) are approximate analytical solutions involving a flux rope and a background magnetic field in equilibrium. 
Most NLFFF solvers presented so far have well reproduced the LL fields, particularly when the analytic solutions are used as boundary conditions at all six boundaries 
\citep{Schrijver06}. The TD models are more difficult to reconstruct, especially in generating a single flux rope structure, but a few codes have done the job 
well \citep{Valori10, Jiang16}. We have also tested our new code against those analytical models, focusing more on the TD models, which have
much more complex field topology than the LL models. 

In this paper, we present a new method of coronal FFF reconstruction and its test against analytical models. Its application to a solar active region will be given 
in a sequel. 
In Section~\ref{sec:pt}, the poloidal and toroidal representation of magnetic field is expounded. Then, we define the problems to be solved and explain 
our numerical algorithm in Section~\ref{sec:num}. In Section~\ref{sec:result}, we present the tests of our new method for TD models in comparison with other methods. 
Lastly, a discussion and summary are provided in Section~\ref{sec:sum}.

\section{Poloidal-Toroidal Representation of Magnetic Field }
\label{sec:pt}

Let us now consider a domain, which encloses a star, but excludes the star, and  
set up a coordinate system such that the stellar surface is a coordinate 
surface of one coordinate named $\xi$.
For example, if the domain is the space exterior to a spherical star of radius $R$, 
the stellar boundary is the surface $r = R$, i.e., $\xi=r$, and 
the natural choice of the coordinates would be spherical coordinates $(r, \theta, \varphi)$. 
If the domain is a semi-infinite space above a plane,  the planar stellar boundary is the surface $z=0$, i.e., $\xi=z$, 
and 
the natural choice of the coordinates would be Cartesian coordinates 
$(x, y, z)$ or cylindrical coordinates $(\rho, \varphi, z)$. 
A magnetic field (a solenoidal vector field) in such a domain can in general be decomposed into 
a poloidal field ${\bm B}_P$ and a toroidal field ${\bm B}_T$ 
\citep{Elsasser46, Lust54, Chandrasekhar57, Chandrasekhar61, Backus58, Radler74, Stern76, Backus86, Low06, Low15, Berger18, Yi22}, i.e., 
\begin{equation} \label{eq:bpt}
{\bm B} = {\bm B}_P + {\bm B}_T \, ,
\end{equation}
in which 
\begin{align}
\label{eq:bp}
& {\bm B}_P  =  \nabla \times ( \nabla \xi \times \nabla \Phi )\, ,
\\
\label{eq:bt}
& {\bm B}_T = \nabla \xi \times \nabla \Psi  \, .
\end{align}
Here    
$\nabla \xi = {\hat{\bm r}}$ in spherical coordinates and $\nabla \xi = {\hat{\bm z}}$ in 
Cartesian or cylindrical coordinates. 
The scalar fields $\Phi$ and $\Psi$, respectively, are called 
poloidal and toroidal scalar functions \citep{Backus86} or Chandrasekhar-Kendall functions \citep{Montgomery78, Low06}. 
We will simply name them the poloidal and toroidal functions in this paper, and we will call the 
magnetic field description given by equations~(\ref{eq:bpt})--(\ref{eq:bt}) the poloidal-toroidal representation (hereafter PT representation)\footnote{
The PT representation has also been called 
the Mie representation \citep{Backus86} or the Chandrasekhar-Kendall representation 
\citep{Montgomery78, Low06} crediting \citet{Mie08} and \citet{Chandrasekhar57}, respectively. However, the credited works only
addressed solutions of linear vector Helmholtz equations, to which linear FFFs also belong. For such a field, 
$\Phi$ and $\Psi$ are not independent of each other while for a general magnetic field, $\Phi$ and $\Psi$ are independent. 
The existence and uniqueness of $\Phi$ and $\Psi$ for an arbitrary $\bm B$ was first treated by 
\citet{Backus58}, and Chandrasekhar's description of general magnetic fields by two independent scalar functions
was first given in his single-authored book \citep{Chandrasekhar61}. 
}.
As can be seen in equations~(\ref{eq:bp}) and (\ref{eq:bt}),  
${\bm B}_P$ and ${\bm B}_T$ are individually divergence-free (solenoidal), and so is $\bm B$ naturally. 
This is one of the merits of the PT representation when used in numerical computation. 
Even if a numerical expression of ${\bm B}$ is not exactly divergence-free, the discretization error does not increase with time (iterations) nor
accumulate in some places. Thus, we are relieved of the concern about $\nabla \cdot {\bm B}$.

We can decompose a magnetic field ${\bm B} ({\bm r})$ into two parts as  
\begin{equation} \label{eq:bdecomp}
{\bm B} ({\bm r}) = {\bm B}_t ({\bm r}) + {\bm B}_n ({\bm r}) \, ,
\end{equation}
in which ${\bm B}_n ({\bm r}) = {\hat{\bm n}} ({\hat{\bm n}} \cdot {\bm B} )$, where ${\hat{\bm n} = \nabla \xi / |\nabla \xi |}  $,
is the component of ${\bm B} ({\bm r})$ normal to 
a constant-$\xi$ surface containing the point ${\bm r}$, and ${\bm B}_t ({\bm r}) ={\bm B} - {\hat{\bm n}} ({\hat{\bm n}} \cdot {\bm B})$ is the tangential 
component. 
%
%
Equation (\ref{eq:bt}) tells that ${\bm B}_T$ has only tangential components to the constant-$\xi$ surfaces, i.e., 
\begin{equation} \label{eq:btn0}
{\hat{\bm n}} \cdot {\bm B}_T   = 0  \, ,
\end{equation}
and any field line of  ${\bm B}_T$ entirely lies in a coordinate surface of $\xi$.
On the other hand, 
${\bm B}_P$ has both tangential and normal components in general, and 
\begin{equation} \label{eq:bnbpn}
B_n = {\hat{\bm n}} \cdot  {\bm B}  = {\hat{\bm n}}  \cdot {\bm B}_P  \, . 
\end{equation}
This property is valid for an arbitrary scalar field $\xi$.  
However, not all formulations in the form of equations 
(\ref{eq:bpt})--(\ref{eq:bt}) with an arbitrary $\xi$ are qualified to be called a standard PT representation, 
which additionally demands that  the curl of a poloidal field be a toroidal field as 
the curl of a toroidal field is a poloidal field, i.e., 
\begin{equation}
\label{eq:curlbp}
  \nabla \times {\bf B}_P = \nabla \xi \times \nabla \Theta \, ,
\end{equation} 
where $\Theta$ is another scalar field. 
This requirement is met when the constant-$\xi$ surfaces are either parallel planes or concentric spheres
\citep{Radler74, Yi22}. In other words, it is required that $\xi= f(r)$, a function of $r$ only in spherical coordinates, 
or $\xi= f(z)$, a function of $z$ only in Cartesian and cylindrical coordinates. 
Fortunately, stars, the sun included, are almost perfectly of a spherical shape, and the base of an individual active region can well be 
approximated as a plane. 
In a standard PT representation, equation~(\ref{eq:curlbp}) tells that
\begin{equation} \label{eq:jpn0}
{\hat{\bm n}} \cdot \nabla \times  {\bm B}_P  = 0  \, , 
\end{equation}
and it follows that 
\begin{equation} \label{eq:jncbt}
J_n= {\hat{\bm n}} \cdot \nabla \times  {\bm B} = {\hat{\bm n}} \cdot \nabla \times  {\bm B}_T  \, .
\end{equation}

In the coronal FFF reconstruction, what is good about equations (\ref{eq:bnbpn}) and (\ref{eq:jncbt}) is 
that $B_n$ and $J_n$ can be fully described by 
the functions $\Phi$ and $\Psi$ in a constant-$\xi$ surface only. 
In a Cartesian coordinate system, 
\begin{align}
\label{eq:bnc}
& B_n = B_z = \nabla_{xy}^2 \Phi \, ,
\\
\label{eq:jnc}
& J_n = J_z = \nabla_{xy}^2 \Psi \, ,
\end{align}
where 
\begin{equation} \label{eq:del2xy}
\nabla_{xy}^2  
= { { {\partial^2} } \over { {\partial} {x^2} } } + { { {\partial^2} } \over { {\partial} {y^2} } } 
\end{equation}
is the 2D Laplacian operator in a $z=const.$ plane. 
In a spherical coordinate system, 
\begin{align}
\label{eq:bns}
& B_n = B_r =  \nabla_{\theta \varphi}^2 \Phi  \, ,
\\
\label{eq:jns}
& J_n = J_r = \nabla_{\theta \varphi}^2 \Psi \, ,
\end{align}
where 
\begin{equation} \label{eq:del2tp}
\nabla_{\theta \varphi}^2  
=  { 1 \over { r^2 {\sin \theta  } } }   { \partial \over { \partial \theta} } \left( \sin \theta { {\partial} \over {\partial \theta} } \right) 
+ { 1 \over { r^2 {\sin^2  \theta  } } }  { { \partial^2} \over { \partial \varphi^2} }
\end{equation}
is the 2D Laplacian operator in an $r = const. $ surface (sphere). 
Since the forms of $\nabla_{xy}^2$ and $\nabla_{\theta \varphi}^2$ do not include any normal derivatives, $B_n$ and $J_n$ are 
fully described by the boundary values of $\Phi$ and $\Psi$ only.  
Thus, implementing the boundary conditions $B_n$ and $J_n$ in a numerical grid is quite straightforward.

In this paper, we will confine ourselves to rectangular domains with Cartesian coordinates. 
In the vector potential description of magnetic field, 
\begin{align}
\label{eq:jnvp}
J_z  & = { \partial \over {\partial z} } \left( \nabla \cdot {\bm A} \right) - \nabla^2 A_z \nonumber \\
     & = { \partial \over {\partial z} } \left( \nabla_{xy} \cdot {\bm A}_{xy} \right) - \nabla_{xy}^2 A_z \, ,
\end{align}
where $\displaystyle \nabla_{xy} = {\hat{\bm x}} {\partial \over {\partial x} } + {\hat{\bm y}} {\partial \over {\partial y} }$ is the 2D $\nabla$-operator 
and ${\bm A}_{xy} = A_x {\hat{\bm x}} + A_y {\hat{\bm y}}$. 
The above equation shows that at least first order normal derivatives of two components of ${\bm A}$ are  
required for describing $J_z (z=0) $. Thus, the boundary values of some components of ${\bm A}$ 
must be changed according to the variation of ${\bm A}$ inside the domain.  This laborious adjustment of 
${\bm A} (z=0)$ can be avoided when the gauge
\begin{equation}
\label{eq:gauge}
\nabla_{xy} \cdot {\bm A}_{xy} = 0
\end{equation}
is employed, not only at the boundary $z=0$, but also inside the domain. 
The gauge (\ref{eq:gauge}) is satisfied when 
\begin{equation}
\label{eq:axyphi}
{\bm A}_{xy} ({\bm r}) = {\hat{\bm z}} \times \nabla \Phi ({\bm r}) \, ,
\end{equation}
where $\Phi $ is an arbitrary scalar field defined in the domain. 
Thus, the vector potential under this gauge is in the form 
\begin{equation}
\label{eq:Agauge}
{\bm A} = {\hat{\bm z}} \times \nabla \Phi  + A_z {\hat{\bm z}} \, ,
\end{equation}
and the resultant magnetic field is in the form
\begin{equation} \label{eq:Bgauge}
{\bm B} =\nabla \times \left( {\hat{\bm z}} \times \nabla \Phi \right) + \nabla \times A_z {\hat{\bm z}} \, .
\end{equation}
Comparing the above equation with equations (\ref{eq:bpt})--(\ref{eq:bt}), we can see that 
the $\Phi$ in equation~(\ref{eq:Bgauge}) is nothing but the poloidal function $\Phi$ in 
equation~(\ref{eq:bp}) and the $A_z$ is $-\Psi$ in equation~(\ref{eq:bt}). 
In this paper, the $\Phi$ above is our poloidal function, but for our toroidal function,  $A_z$ is taken instead of $- \Psi$. 
Therefore,  we have the following expressions for the respective three components of ${\bm B}$ and ${\bm J}$, 
to be used for our numerical computation. 
\begin{align}
\label{eq:bx}
& B_x = 
- { \partial \over {\partial x} } \left( { {\partial \Phi } \over {\partial z }  } \right) + { {\partial A_z } \over {\partial y } } 
\, ,
\\
\label{eq:by}
& B_y = 
- { \partial \over {\partial y} } \left( { {\partial \Phi } \over {\partial z }  } \right) - { {\partial A_z } \over {\partial x } } 
\, ,
\\
\label{eq:bz}
& B_z = 
{ { {\partial^2} {\Phi} } \over { {\partial} {x^2} } } + { { {\partial^2} {\Phi} } \over { {\partial} {y^2} } } 
 \, .
\\
\label{eq:jx}
& J_x = 
{ \partial \over {\partial x} } \left( { {\partial A_z } \over {\partial z }  } \right) 
+ { \partial \over {\partial y} } \left( \nabla^2 \Phi \right) 
 \, ,
\\
\label{eq:jy}
& J_y = 
{ \partial \over {\partial y} } \left( { {\partial A_z } \over {\partial z }  } \right) 
- { \partial \over {\partial x} } \left( \nabla^2 \Phi  \right) 
 \, ,
 \\
\label{eq:jz}
& J_z = 
 - \left ( { { {\partial^2} {A_z} } \over { {\partial} {x^2} } } + { { {\partial^2} {A_z} } \over { {\partial} {y^2} } } \right)
 \, .
\end{align}
The first terms in the righthand side of equations (\ref{eq:bx}) and (\ref{eq:by}) are  
the 2D curl-free (irrotational) part of the tangential magnetic field 
$\displaystyle {\bm B}_t = {\bm B}_{xy} = B_x {\hat{\bm x}} + B_y {\hat{\bm y}} $,
and the second terms form its 2D divergence-free (solenoidal) part.
The same is true for ${\bm J}_t = {\bm J}_{xy} = J_x {\hat{\bm x}} + J_y {\hat{\bm y}}$ given by equations (\ref{eq:jx}) and (\ref{eq:jy}). 

A remark should be made that the PT representation 
in a Cartesian coordinate system demands a special treatment when the magnetic field is periodic in $x$ and $y$  \citep{Schmitt92}. 
In our FFF calculations of this paper, we do not employ a periodic boundary condition and thus are relieved of such a special care. 
See Appendix~\ref{app:ptperiod} for some details. 

Before the present paper, 
the PT representation has been used in construction of force-free fields or magnetohydrostatic (MHS) equilibria. 
However, these works are related to 
LFFFs or a special class of MHS equilibria. 
For LFFFs, the toroidal function and poloidal functions are related by $\Psi = \alpha \Phi$, where 
$\alpha=const.$ and the poloidal function is 
a solution of a linear Helmholtz equation 
$\displaystyle \nabla^2 \Phi + \alpha^2 \Phi =0$. This linear problem was treated by an eigenfunction expansion
\citep{Chandrasekhar57, Nakagawa72} and by constructing Green's function based on eigenfunctions \citep{Chiu77}. 
An MHS equilibrium is  
described by 
\begin{equation*}
\label{eq:mhsequil}
{\bm J} \times {\bm B} - \nabla p - \rho \nabla \psi = 0 \, ,
\end{equation*}
where $p$ is the plasma pressure and $\psi$ the gravitational potential.  
One can then express the current density in 
the form of  
\begin{equation*}
\label{eq:mhsequil2}
{\bm J} = {\tilde \alpha} {\bm B} + \nabla \mu \times \nabla \psi \, ,
\end{equation*}
in which 
$\displaystyle 
{\tilde \alpha} = ( {\bm J} \cdot \nabla \psi  ) / ( {\bm B} \cdot \nabla \psi ) $ and $\displaystyle \mu$ is a function of ${\bm B}\cdot \nabla \psi$ and  $\psi$ 
\citep[see][for details]{Low91}.  
In general MHS equilibria, it holds that ${\bm B}\cdot \nabla {\tilde \alpha}=0$, i.e., ${\tilde \alpha}$ is a function of each field line. 
In a special class of MHS equilibria, in which 
$\displaystyle 
{\tilde \alpha} = const. $ everywhere, 
it holds that $\Psi = {\tilde \alpha} \Phi$, and the poloidal function $\Phi$ satisfies an equation 
$\displaystyle 
\nabla^2 \Phi + {\tilde \alpha}^2 \Phi - {\mathcal F } = 0$, where  ${\mathcal F }$ is a certain function of $\mu$
\citep{Neukirch99b}. This problem was solved with a Green's function method with an eigenfunction decomposition by \citet{Petrie00}. 
Thus, all the previous works employing the PT representation have solved linear problems for the poloidal function only.  
Construction of NLFFFs employing independent poloidal and toroidal functions was proposed by \citet{Neukirch99a}, but
has not been worked out in detail.

\bigskip
 
\section{Numerical Algorithm and Modeling} 
\label{sec:num}

\subsection{Numerical Algorithm} \label{subsec:numal}

As pointed out by \citet{Grad58}, an FFF problem given by equations~(\ref{eq:fff}) and (\ref{eq:gauss}) is semi-elliptic and semi-hyperbolic. 
The hyperbolic nature of the problem lies in equation~(\ref{eq:alpha}), which is called a magnetic differential equation \citep{Kruskal58}. 
The Grad-Rubin type procedure includes a solver of the magnetic differential equation, practically putting more weight on 
the hyperbolic nature of the problem. In variational methods, the elliptic nature is more emphasized and the hyperbolic nature is rather
implicitly considered. Our new algorithm is more inclined to variational methods in that we solve elliptic equations at every iteration step with 
the hyperbolic nature of the problem considered in the source terms. However, it is not variational because we do not try to extremize any functional. 
The rationale and details of our new method are given below. 

Let us simply consider a magnetofrictional evolution of magnetic field under ideal MHD condition. 
Then, the electric field is given by
$\displaystyle {\bm E} = - {\bm v} \times {\bm B}$ after a proper normalization removing constants involved in unit systems. 
We can set 
\begin{equation}
\label{eq:vnjb}
{\bm v} = \nu ( {\bm r}, t ) \, { { {\bm J} \times {\bm B}  } \over {B^2} } \, ,
\end{equation}
where $\nu  ( {\bm r}, t ) $ is an arbitrary scalar field, which may be set to unity in nondimensionalized equations.  
Then
\begin{align}
{\bm E} & = -{ { ( {\bm J} \times {\bm B}  ) \times {\bm B} } \over {B^2} }
= {\bm J} - {\hat{\bm b}} ( {\hat{\bm b}} \cdot {\bm J} ) \nonumber \\
& =  {\bm J} - {\bm J}_\parallel  = {\bm J}_\perp \, ,
\label{eq:ejperp}
\end{align}
in which ${\hat{\bm b}} = {\bm B} / B$ is a unit vector in the direction of $\bm B$, and 
${\bm J}_\parallel = {\hat{\bm b}} ( {\hat{\bm b}} \cdot {\bm J} ) $ and 
${\bm J}_\perp = {\bm J} - {\bm J}_\parallel $ are respectively component vectors of ${\bm J}$ parallel and perpendicular to ${\bm B}$. 
Therefore, the induction equation (Faraday's law) reads
\begin{equation}
\label{eq:dbdt}
 { {\partial {\bm B}} \over {\partial t} } = -\nabla \times {\bm J}_\perp = - \nabla \times ( {\bm J} -  {\bm J}_\parallel  ) \, .
\end{equation}
Since the righthand side of this equation involves second order spatial derivatives of $\bm B$, it is a parabolic equation. 
As $t \rightarrow \infty$, both sides of the parabolic equation go to zero. The asymptotic state $\nabla \times {\bm J}_\perp = 0$ means 
that  ${\bm J}_\perp =  \nabla \phi$, where $\phi$ is a certain scalar field. If the field line footpoints at the boundary are fixed, ${\bm E}=0$ at the boundary and 
$\nabla \phi = 0$ everywhere in the domain. In the coronal FFF problems, the stressing (winding) and the relaxing (unwinding) are alternating or they indistinguishably coexist. 
In any case, as $t \rightarrow \infty$, the 
stressing is reduced to zero, and the asymptotic state is such that ${\bm J}_\perp = 0$, i.e., $ {\bm J} = {\bm J}_\parallel $,  a force-free state. 

Now the PT representation has two variables describing magnetic field, $\Phi$ and $A_z$ ($=  -\Psi$), which are directly related to 
$B_z$ and $J_z$ by equations~(\ref{eq:bz}) and (\ref{eq:jz}) respectively. 
From equation~(\ref{eq:dbdt}), we can write down the evolutionary equation of $B_z$ and $J_z$ as 
\begin{align}
 { {\partial {B_z}} \over {\partial t} } 
 & = \nabla^2 B_z +  \left(  { {\partial {J_{\parallel y}}} \over {\partial x} }  - { {\partial {J_{\parallel x}}} \over {\partial y} }   \right) \ ,
 \label{eq:dbzdt}
 \\
 { {\partial {J_z}} \over {\partial t} } 
& = { \partial \over {\partial z} } \nabla \cdot {\bm J}_\parallel - \nabla^2 {J_{\parallel z}} + \nabla^2 J_z \, , 
 \label{eq:djzdt} 
\end{align}
where $t$ is a sort of ``pseudo-time.'' 
Setting the righthand sides of these equations to zero ($t = \infty$) gives the following force-free conditions:  
\begin{align}
\nabla^2 B_z & =    { {\partial {J_{\parallel x}}} \over {\partial y} }  - { {\partial {J_{\parallel y}}} \over {\partial x} }   \, ,
 \label{eq:bzffc}
 \\
  \nabla^2 J_z
& =  \nabla^2 {J_{\parallel z}}  - { \partial \over {\partial z} } \nabla \cdot {\bm J}_\parallel  \nonumber \\
& = \nabla_{xy}^2 {J_{\parallel z}} - { \partial \over {\partial z} } \nabla_{xy} \cdot {\bm J}_\parallel \, .
 \label{eq:jzffc} 
\end{align}
It would be more than desirable to solve for $B_z$ and $J_z$ at once, but the righthand sides of the above equations are
so highly nonlinear as to make such an attempt appear hopelessly difficult. The lefthand sides of the equations are linear and 
purely elliptic while the hyperbolic nature of the FFF equations is contained in the righthand sides. 
With this understanding, we propose the following iteration procedure: 
\begin{align}
\nabla^2 {B_z}^{m+1} & =    { {\partial {J_{\parallel x}}^m} \over {\partial y} }  - { {\partial {J_{\parallel y}}^m} \over {\partial x} }   \, ,
 \label{eq:bziter}
 \\
 \nabla_{xy}^2 \Phi^{m+1} & = {B_z}^{m+1}   \, ,
 \label{eq:phiupd}
 \\
  \nabla^2 {J_z}^{m+1}
& =  \nabla^2 {J_{\parallel z}}^m  - { \partial \over {\partial z} } \nabla \cdot { {\bm J}_\parallel }^m  \nonumber \\
& = \nabla_{xy}^2 {J_{\parallel z}}^m - { \partial \over {\partial z} } \nabla_{xy} \cdot { {\bm J}_\parallel }^m \, , 
 \label{eq:jziter} 
 \\
  \nabla_{xy}^2 {A_z}^{m+1} & = - {J_z}^{m+1} \, , 
 \label{eq:azupd} 
\end{align}
in which the superscript denotes the iteration step. Here we are to solve simple 3D Poisson equations for $B_z$ and $J_z$ with source terms 
evaluated with known values. Then, the poloidal function $\Phi$ and the toroidal function $A_z$ are updated by solving 2D 
Poisson equations at each $z=const.$ plane, 
and all the other variables at the $(m+1)$-th step are calculated from $\Phi^{m+1}$ and ${A_z}^{m+1}$ 
by equations~(\ref{eq:bx}), (\ref{eq:by}), (\ref{eq:jx}) and (\ref{eq:jy}). 
The iteration procedure given by equations~(\ref{eq:bziter})--(\ref{eq:azupd}) is 
not variational because it does not try to extremize any functional. The iteration step $m$ is far from representing any pseudo-time because 
we are already at  $t = \infty$. 
Although we currently cannot present a mathematical proof of the convergence of our algorithm,  
our tests with many different real and artificial magnetograms have never failed in convergence. 
A similar iterative method was used for solving 2D steady-state Navier-Stokes equations described in stream function and vorticity \citep{Roache75}.

\subsection{Boundary and Initial Conditions} \label{subsec:bndini}

In this paper, we consider a rectangular domain $V =\{ (x,y,z) | {0 \le x \le L_x,} \ {0 \le y \le L_y,} \ {0 \le z \le L_z} \}$\footnote{
A domain $V =\{ (x,y,z) | {x_a \le x \le x_b,} \ {y_a \le y \le y_b,} \ {z_a \le z \le z_b} \}$ can always be expressed in that form by 
a displacement of the origin. 
}.
Its boundary consists of six planes. The $z=0$ plane 
corresponds to the coronal base, and the $z=L_z$ plane is an artificial top boundary of the corona. The lateral boundary of an active region corona 
consists of four planes: $x=0$, $x=L_x$, $y=0$ and $y=L_y$. A numerical grid is set up in the computational domain such that 
$G = \{ (i, j, k) |\, i = 0, 1, 2, \ldots, N_x, \ j = 0, 1, 2, \ldots, N_y, \ k = 0, 1, 2, \ldots, N_z \}$. 

At the bottom boundary ($z=0$), $B_z$ and $J_z$ are given, which are readily translated into $\Phi$ and $A_z$ by solving 2D Poisson equations 
$\displaystyle \nabla_{xy}^2 \Phi = {B_z} $ and $\displaystyle \nabla_{xy}^2 A_z = -{J_z} $ once for all. Although imposing $B_z$ and $J_z$  
at $z=0$ is straightforward owing to the PT representation of magnetic field, evaluating the righthand side of 
equation~(\ref{eq:jziter}) at the grid plane $k=1$ is somewhat tricky because the finite difference form of the $z$-derivative term
$\displaystyle { \partial \over {\partial z} } \nabla_{xy} \cdot { {\bm J}_\parallel }$ at $k=1$ (the first grid plane above the bottom boundary) 
requires the value of ${ {\bm J}_\parallel }$ at $k=0$ ($z=0$). 
As can be seen in equations~(\ref{eq:jx}) and (\ref{eq:jy}), $J_x$ and
$J_y$ involve a second order $z$-derivative $\partial^2 \Phi / \partial z^2$, which cannot unambiguously be defined at $z=0$. 
However, the term $ { {\bm J}_\parallel } (z=0) $ does not need to be evaluated at all. Since the equations we are solving (eqs.~[\ref{eq:bzffc}] and [\ref{eq:jzffc}])
already represent a stationary state ($t = \infty$), we should rather set $ {\bm J}_\perp (z=0) = 0 $. 
Using the relation $\displaystyle  \nabla \cdot { {\bm J}_\parallel } = - \nabla \cdot { {\bm J}_\perp }$, we can rewrite equation~(\ref{eq:jzffc}) as 
\begin{align}
\nabla^2 J_z
& =  \nabla^2 {J_{\parallel z}}  + { \partial \over {\partial z} } \nabla \cdot {\bm J}_\perp  \nonumber \\
& = \nabla_{xy}^2 {J_{\parallel z}} + { \partial \over {\partial z} } \nabla_{xy} \cdot {\bm J}_{\perp xy} 
+ { {\partial^2 J_z} \over  { \partial z^2} } \, .
 \label{eq:jzffc2} 
\end{align}
The stationary condition ${\bm J}_\perp (z=0) = 0$ allows us to evaluate the second term in the rightmost hand side of the above equation straightforwardly. 
The implementation of the boundary condition at $z=0$ is thus completed. 
We have tried another method, which uses a plausible relation 
\begin{equation}
{\bm J}_{\parallel xy} (x, y, z=0) = \alpha (x, y, 0) \, {\bm B}_{xy}^\ast (x, y, 0) \, ,
\label{eq:jalphab}
\end{equation}
in which $\alpha = J_z / B_z$ and ${\bm B}_{xy}^\ast $ is the observed horizontal magnetic field. 
Using this relation corresponds to employing the observed three components of magnetic field at $z=0$ and 
enforcing the force-free condition at the bottom boundary. 
The results of both methods are compared for many cases and they turn out to be surprisingly similar to each other. 

The magnetogram of an active region generally does not have the same positive and negative magnetic fluxes because some field lines 
are connected to the outside of the active region as closed fields or to the solar wind as open fields. We want to accommodate 
such a flux imbalance in the new FFF construction method. Previous attempts to consider a flux imbalance have placed image polarities of 
opposite signs outside the real domain \citep{Seehafer78, Otto07}. 
These methods, however, may make a considerable flux go out of or come into the real 
domain through the lateral boundary and thus the constructed field configuration may undesirably depend on the field connection between the real domain and 
the image domains rather than on the field connection within the real domain alone. 
In our study, we rather confine all the flux within the computational domain except the flux passing through the top boundary. 
Magnetic field lines thus cannot penetrate the lateral boundary, but can be tangential to it. 
The effects of other boundary conditions allowing nonzero magnetic flux through the lateral boundary are discussed in Section~\ref{sec:sum}.

In regard to the lateral boundary ($x=0, L_x$,  $y=0, L_y$),  the unit outward normal vector to the boundary is denoted by $\hat{\bm n}$,
 the normal component vector of a vector field 
$\displaystyle {\bm F}$ by ${\bm F}_n =  {\hat{\bm n}} ( {\hat{\bm n}} \cdot {\bm F} )$, its tangential component vector by
$\displaystyle {\bm F}_\parallel = {\hat{\bm n}} \times ( {\bm F} \times {\hat{\bm n}}) = {\bm F} - {\bm F}_n$, and the normal derivative by 
$\displaystyle \partial / \partial n = {\hat{\bm n}} \cdot \nabla$. Since we have to solve four Poisson equations, i.e., equations~(\ref{eq:bziter})--(\ref{eq:azupd}), at every
iteration step, we have set up four boundary conditions respectively on $B_z$, $\Phi$, $J_z$ and $A_z$ at the lateral boundary. 
To make this boundary impenetrable to magnetic field, i.e., $B_n=0$, we choose the following boundary conditions. 
\begin{align}
{\partial B_z} \over {\partial n} & = 0 \, , 
\label{eq:bzbc}
\\
{\partial \Phi} \over {\partial n} & = p_1 = const. \, ,
\label{eq:phibc}
\\
J_z & = 0 \, ,
\label{eq:jzbc}
\\
A_z & = 0 \, .
\label{eq:azbc}
\end{align}
As can be seen in equations~(\ref{eq:bx}) and (\ref{eq:by}), the conditions~(\ref{eq:phibc}) and (\ref{eq:azbc}) result in $B_n = 0$. 
The 2D Poisson equation~(\ref{eq:phiupd}) gives the following constraint: 
\begin{equation}
\oint {{\partial \Phi} \over {\partial n}} dl = \int B_z dS \, , 
\label{eq:phiconst}
\end{equation}
in which $dl$ is a unsigned line element surrounding the 2D domain at $z=const.$ and $dS$ is the area element in the plane. 
Then, $p_1$ in equation~(\ref{eq:phibc}) is given by
\begin{equation}
p_1 = { 1 \over { \left( 2 L_x + 2 L_y \right) } } \,  {\int_0^{L_y}  \int_0^{L_x}  B_z \, dx \, dy    }  \, ,
\label{eq:p1}
\end{equation}
If an infinite plane is a perfectly conducting rigid wall impenetrable to magnetic field, the magnetic field in one side of the wall can be regenerated by replacing the wall by 
an image field beyond the wall in such a way that ${\bm B}_t$ is symmetric and $B_n$ is anti-symmetric across the boundary. In our case, the lateral boundary consists 
of four finite planes and the magnetic flux in each $z=const.$ plane is unmatched, which does not realize an exact symmetry or anti-symmetry across each boundary plane. 
Nevertheless, the local symmetry condition~(\ref{eq:bzbc}) 
turns out to be in practice a good choice for our situation too. 
The image field beyond an infinite wall also gives ${\bm J}_t =0$ and $\partial J_n / \partial n = 0$. 
The boundary condition~(\ref{eq:jzbc}) is also inspired by this picture. 

At the top boundary, we employ a source surface condition \citep{Altschuler69, Aly93}. The purpose of this boundary condition is not only to mimic the 
upper corona, where the solar wind carries out the open magnetic field, but also to provide an exit for the imbalanced magnetic flux. The source surface boundary condition 
is simply given by
\begin{align}
\label{eq:bxbyss}
& B_x (z=L_z) = B_y (z=L_z) = 0 \, ,
\\
\label{eq:bzss}
& { {\partial B_z} \over {\partial z} } \Big|_{z=L_z} = 0 \, .
\end{align}
The condition~(\ref{eq:bxbyss}) results in 
\begin{equation}
\label{eq:jzss}
J_z (z=L_z) = 0 \, ,
\end{equation}
implying that no current can escape or come in through the top boundary. Since $J_n=0$ at the lateral boundary in the final force-free state, 
it follows that
\begin{equation}
\label{eq:Jzfluxb}
\int_{z=0} J_z \, dx \, dy = 0 \, ,
\end{equation}
i.e., the positive and negative currents through the bottom boundary should be balanced.
This constraint must be fulfilled by preprocessing of the magnetogram data. 
The source surface condition at the top boundary is known to create a magnetic field null-line at the boundary and a current sheet connected to it in the domain 
\citep{Zwingmann85, Platt94} except for potential fields. 
This point will be discussed in relation to our numerical solutions in Section~\ref{sec:sum}. 

To start the iteration procedure given by equations~(\ref{eq:bziter})--(\ref{eq:azupd}), we need zeroth step ($n=0$) values of 
$B_z$, $\Phi$, $J_z$ and $A_z$, which may be called initial conditions. So far many numerical calculations for FFFs have taken potential fields for 
their initial conditions \citep{Schrijver06}. In the PT representation, a potential field is prescribed by
\begin{subequations}
\begin{align}
\nabla^2 {B_z} & =    0   \, ,
 \label{eq:bzpot}
 \\
 \nabla_{xy}^2 \Phi & = {B_z}   \, ,
 \label{eq:phipot}
 \\
J_z & = 0 \, , 
 \label{eq:jzpot} 
 \\
 A_z & =0 \, ,  
 \label{eq:azpot} 
\end{align}
\end{subequations}
in which the last line results from equation~(\ref{eq:jzpot}) and our lateral boundary condition~(\ref{eq:azbc}). 
Since equations~(\ref{eq:jzpot}) and (\ref{eq:azpot}) do not match our boundary conditions at $z=0$, 
the potential field as an initial condition will have awkward bending of field lines just above the bottom boundary. 
Instead of starting with a current-free state, we rather want to load the entire domain with currents. The simplest way for this 
would be to set 
\begin{equation}
\nabla \times {\bm J}  = \nabla \times \nabla \times {\bm B} = 0 \, . 
\label{eq:curlj0}
\end{equation}
This field is prescribed by
\begin{subequations}
\begin{align}
\nabla^2 {B_z} & =    0   \, ,
 \label{eq:bzcj}
 \\
\nabla^2 {J_z} & = 0 \, , 
 \label{eq:jzcj} 
\end{align}
\end{subequations}
with 
$\displaystyle \nabla_{xy}^2 \Phi = {B_z}$ and $\displaystyle \nabla_{xy}^2 A_z =- {J_z}$. 
In our experience, this initial field reduces the number of iteration steps required for convergence compared with 
the potential field as an initial condition.

\bigskip

\section{Tests of Our New Method against the Titov-D\'{e}moulin Models} 
\label{sec:result}

\subsection{Flux Ropes and Bald Patches of Titov-D\'{e}moulin Models} \label{subsec:td12}

\begin{deluxetable*}{ccccccccc}
\tablenum{1}
\tablecaption{Parameters Prescribing Three TD models \label{tb:tdpara}}
\tablewidth{0pt}
\tablehead{
\colhead{Model}&
\colhead{$R \,  [10^6\, {\mathrm m}] $}&
\colhead{$a \,  [10^6\, {\mathrm m}] $} &
\colhead{$L \,  [10^6\, {\mathrm m}] $} &
\colhead{$d \,  [10^6\, {\mathrm m}] $} &
\colhead{$q \, [10^{12}\, {\mathrm {Wb}}\, ] $}&
\colhead{$I_0 \, [10^{12}\, {\mathrm A}] $}&
\colhead{$N_t  \, [ {\mathrm {turns} } ] $\tablenotemark{a}}&
\colhead{No. of grid-points}
}
\startdata
TD1 \tablenotemark{b} & $110$ & $35$ & $100$ & $50$ & $100$ & $-13$ & $2.577 $ & $151 \times 251 \times 101$  \\
TD2 \tablenotemark{c} & $100$ & $35$ & $60$ & $60$ & $80$ & $-3$ & $8.8$ & $151 \times 251 \times 101$  \\
TD3 \tablenotemark{d} & $110$ & $32.5$ & $50$ & $50$ & $100$ & $-7$ & $5.0$  & $257 \times 351 \times 201$ \\
\enddata

\tablenotetext{a}{The field line twist at the surface of the entire closed torus, which is larger than the twist of the coronal part of the torus.}
\tablenotetext{b}{Used in \citet{Wiegelmann06b}. }
\tablenotetext{c}{Case 5 of \citet{Demcsak20}. }
\tablenotetext{d}{Used in \citet{Torok04}. }
\end{deluxetable*}

As with other numerical methods, our method is to be tested with a computational code built upon it against known analytical solutions. 
So far most existing codes have been tested against the analytical model by \citet{Low90}, which represents a class of moderately sheared magnetic fields, and have shown 
good to excellent performances \citep{Schrijver06}.   
Our own variational code VFVP ({\bf V}ariational {\bf F}FF Code in {\bf V}ector {\bf P}otential Formulation) we developed earlier (see Appendix~\ref{app:var}) and 
the code based on our new method 
NFPT ({\bf N}on-variational {\bf F}FF Code in {\bf P}oloidal-{\bf T}oroidal Formulation) also are found to be working as well for the Low \& Lou model as other codes are \citep{Schrijver06, Inoue14}, 
and we do not feel the necessity of presenting the results here. In this paper, we present the performance of the codes in reproducing the NLFFF model by \citet{Titov99}. 
In this analytical model, a flux rope carrying a helical force-free field and ambient fields generated by two magnetic charges and a line current underneath the surface are 
in equilibrium. Each Titov-D\'{e}moulin (TD) model is prescribed by six free parameters: $R$ the major radius of the torus, $a$ the minor radius, 
$L$ the half-distance between two magnetic charges, $q$ the magnitude of the magnetic charge, $d$  the distance between the photosphere and 
the subsurface line current lying along the symmetry axis of the torus, and $I_0$ the subsurface line current. 
The toroidal current $I$ in the torus is determined by the force balance condition. 
The field line twist number $N_t$ at the surface of the entire (closed) torus also comes out from these parameters.  
The coronal part of the torus, which lies above the photospheric plane, has a twist $N_{\mathrm {cor}}$ at its surface 
\begin{equation} \label{eq:ntncor}
N_{\mathrm {cor}} \approx { {N_t}\over {\pi } } \cos^{-1} \left( {d \over R}   \right) \, .
\end{equation}
Table~\ref{tb:tdpara} lists the parameters prescribing the three TD models to be presented in this paper. 
It is to be noted that since the TD model employs a thin flux tube approximation assuming a high aspect ratio ($A=R/a \gg 1$), 
the minor radius of the flux tube may not be uniform along its axis in numerical models with a modest $A$, 
particularly when lateral and top boundary conditions are differently prescribed from those of the analytical models.

\begin{figure*}[ht!]
\begin{center}
\includegraphics[width=12cm] {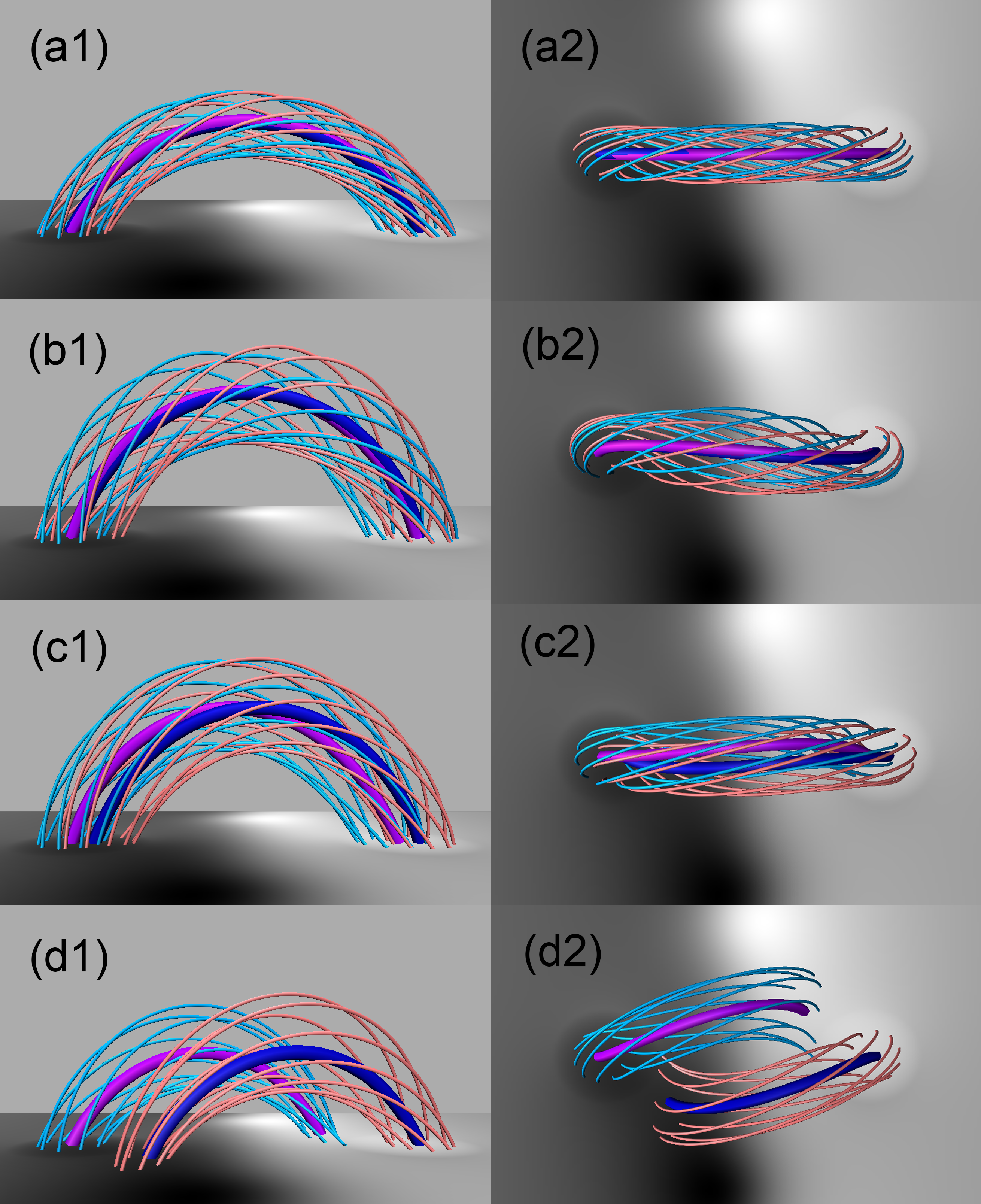}
\end{center}
\caption{Field lines of model TD1. Each row shows a side view and a top view for (a) the analytical model,  
(b) the numerical solution by our new code NFPT (non-variational FFF code in poloidal-toroidal formulation), 
(c) the numerical solution by our earlier code VFVP (variational FFF code in vector potential formulation), and 
(d) the numerical solution by the optimization code in SolarSoftWare (SSW). Field lines in all models are traced from the same footpoints.  
The black and white brightness in the photosphere represents the polarity of the line-of-sight magnetic field component, i.e., white for positive and 
black for negative. 
The thin field lines in magenta are traced from certain positive footpoints of the analytical flux rope and those in cyan from negative footpoints. 
The thick field lines are drawn from the two footpoints
of the magnetic axis (blue from the positive and pink from the negative) of the analytical flux rope. 
For the analytical model (a) and the numerical model (b), the thick field lines overlap with each other. 
For the numerical model (c), they are slightly off, but the overall structure shows one flux rope. For the numerical model (d), 
thick field lines are quite separated and the overall structure indicates the presence of two flux tubes rather than one. 
\label{fig:td1}}
\end{figure*}

Comparison of numerical results and reference models can first be made by their appearance. In each TD model, there is only one flux rope involved. 
It is thus a touchstone of different numerical models whether one flux rope is well reproduced with the characteristic features of the analytical model. 
Figure~\ref{fig:td1} shows the field lines of the flux rope in model TD1 (refer to Table~\ref{tb:tdpara}) obtained from (a) the analytical model, (b) our new code NFPT, (c) our earlier code VFVP, 
and (d) the optimization code in the SolarSoftWare (SSW), which is available in the public domain\footnote{\url{http://sprg.ssl.berkeley.edu/~jimm/fff/optimization_fff.html}}.
All field lines are traced from the same footpoints, one group in the positive polarity area and the other in the negative polarity area of the flux rope in the analytical model. 
The field lines traced from
the positive and negative footpoints of the magnetic axis of the flux rope in the analytical model are rendered thicker than other field lines. The field lines of NFPT and VFVP 
apparently express a single flux rope while those of SSW show two separated flux tubes. The two magnetic axis field lines obtained from our new code NFPT overlap 
with each other
as in the analytical model, but those from our earlier code VFVP are slightly off if not as much as those from SSW. 
Since we do not have observational data at the lateral and top boundaries in practical situations, 
all numerical computations here use the data of the analytical model at the bottom boundary only, 
and proper artificial boundary conditions are used for the lateral and top boundaries. 
For the same TD model, \citet{Wiegelmann06b} also reported two separated flux tubes crossing each other when the analytical solution is used only at the bottom boundary,   
whereas their constructed field bears a remarkable resemblance to the analytical model when the analytical solution is imposed at the lateral and top boundaries too. 

%
%
\begin{deluxetable*}{cccccccc}
\tablenum{2}
\tablecaption{Performance Metrics of Different Solutions for Model TD1 \label{tb:td1metric}}
\tablewidth{0pt}
\tablehead{
\colhead{Solution} &
\colhead{$ C_{\mathrm {vec} } $} &
\colhead{$ C_{\mathrm {CS} } $} &
\colhead{$ E_{\mathrm n}^{'} $} &
\colhead{$ E_{\mathrm m}^{'} $} &
\colhead{$ \epsilon $}&
\colhead{$ \epsilon_p $}&
\colhead{$ CW{\mathrm {sin} } $}
}
\startdata
Analytical  & $1.000$ & $1.000$ & $1.000$ & $1.000$ & $1.000$ & $2.407$ & $0.080$\tablenotemark{a} \\
NFPT & $0.935$ & $0.898$ & $0.590$ & $0.534$ & $0.643$ & $1.573$ & $0.102$   \\
VFVP  & $0.947$ & $0.946$ & $0.671$ & $0.628$ & $0.739$ & $1.781$ & $0.298$   \\
SSW  & $0.920$ & $0.942$ & $0.556$ & $0.472$ & $0.537$ & $1.293$ & $0.662$   
\enddata
\tablenotetext{a}{This value is not zero in a numerical grid.}
\end{deluxetable*}

Beyond mere appearance, we can quantitatively compare numerical models with analytical solutions using so-called ``figures of merit'' devised by \citet{Schrijver06}, 
which will also be called ``performance metrics'' in this paper. Although we will use the same notations for them as in \citet{Schrijver06}, the definitions of those and other metrics 
are reiterated here for readers' convenience. 
\begin{align} 
& C_{\mathrm {vec}}={{\sum\limits_{i}\bm{B}_i\cdot\bm{b}_i} \over {\left(\sum\limits_{i}\left| \bm{B}_i\right|^2 \sum\limits_{i}\left|\bm{b}_{i}\right|^2\right)^{1 \over 2}}}, \label{eq:cvec} \\
& C_{\mathrm {CS} }= {{{1}\over{M}} \sum\limits_{i}{{\bm{B}_{i}\cdot\bm{b}_{i}} \over {\left|\bm{B}_{i}\right|\left|\bm{b}_{i}\right|}}}, \label{eq:ccs} \\
& E_{\mathrm n}^{'}={1-{{\sum\limits_{i}\left|\bm{B}_{i}-\bm{b}_{i}\right|} \over {\sum\limits_{i}\left|\bm{b}_{i}\right|}}}, \label{eq:en} \\
& E_{\mathrm m}^{'}={1-{{1\over M} \sum\limits_{i}{{\left|\bm{B}_{i}-\bm{b}_{i}\right|}\over{\left|\bm{b}_{i}\right|}}}}, \label{eq:em} \\
& \epsilon={{\sum\limits_{i}\left|\bm{B}_{i}\right|^2}\over{\sum\limits_{i}\left|\bm{b}_{i}\right|^2}}, \label{eq:eps} \\
& CW{\mathrm {sin} }=
\displaystyle{\displaystyle{{\sum\limits_{i}{{\left|\bm{J}_{i}\times\bm{B}_{i}\right|}\over{B_{i}}}}}\over{\sum\limits_{i}J_{i}}}, \label{eq:cwsin}
\end{align}
in which $i$ denotes each grid point in the computational domain,  $M$ the total number of grid points, ${\bm b}$ the reference magnetic field, and ${\bm B}$ the numerical solution. 
Ideally, the reference field ${\bm b}$ should be the exact solution under the same boundary conditions used by the numerical solutions ${\bm B}$. However, the boundary conditions 
for ${\bm b}$ and ${\bm B}$ are inevitably different if the lateral and top boundary values of ${\bm b}$ are not known to those who construct ${\bm B}$. 
Also, the analytical exactness is different from 
numerical exactness in a grid of finite resolution. Furthermore, the analytical model of \citet{Titov99} is an approximate solution. For these reasons, some authors have 
\citep[e.g.,][]{Valori10, Guo16}
relaxed the analytical solution in a numerical grid to obtain a reference field. In this paper, however, we use the unprocessed analytical solutions as reference fields 
in spite of the possible 
loss of performance scores. 
The metric $CW{\mathrm {sin}}$ was originally devised by \citet{Wheatland00}, in which a different 
notation $\sigma_J$ was given to it, but 
the current naming has become more popular afterwards.  
While the metrics $C_{\mathrm {vec} }$, $C_{\mathrm {CS}}$, $E_{\mathrm n}^{'}$, $E_{\mathrm m}^{'}$, and 
$ \epsilon$ are based on the comparison of the numerical solution with the reference field, 
the metric $CW{\mathrm {sin}}$ only judges the force-freeness of the numerical solution itself. 
If the reference field and the numerical solution are exactly identical, the former five metrics  should be all $1$. 
For an exact force-free field,  $CW{\mathrm {sin}}$ should be $0$. 
In some literature and also in this paper, another quantity  $\epsilon_p$ is presented, which 
is the magnetic energy of a solution in units of the potential field energy. 
\begin{equation}
\label{eq:epsp}
\epsilon_p={{\sum\limits_{i}\left|\bm{B}_{i}\right|^2}\over{\sum\limits_{i}\left|\bm{B}_{p,i}\right|^2}},
\end{equation}
where ${\bm B}_p$ the potential magnetic field. This metric uses the potential magnetic field as the reference field and indicates the nonpotentiality of the solution, which may 
be due to the deviation of the solution from the exact FFF as well as to the difference of the exact FFF and the potential field. 

Table~\ref{tb:td1metric} lists the seven performance metrics for TD1 derived from the analytical solution and the three numerical 
solutions, NFPT, VFVP and SSW. The presented value of $ CW{\mathrm {sin}} $ for the analytical solution is obtained by assigning the analytical magnetic field to 
a numerical grid and evaluating the current density by a finite-difference method, and hence it deviates from $0$ inevitably.   
As for the first five metrics, the scores are ranked in order of VFVP, NFPT and SSW. However, we have already seen that 
the field obtained by VFVP does 
not show as much resemblance to the analytical solution as that by NFPT. It is only in TD1 that VFVP's scores are better than NFPT's. 
For other TD models, NFPT turns out to exceed VFVP in all metrics, as we shall see further below (Tables 3 and 4).. 
In contrast to the metrics by \citet{Schrijver06}, 
the scores of the metric $ CW{\mathrm {sin}} $ are ranked in order of 
NFPT, VFVP and SSW, reflecting best the resemblance of the numerical solutions to the TD field featured by one flux rope.  
The metric $ CW{\mathrm {sin}} $ is independent of the choice of the lateral and top boundary conditions and evaluates the force-freeness 
of a solution itself. It is interesting that the apparent resemblance depends more on the exactness (force-freeness) of the solution than its 
numerical proximity to a reference solution. In other TD models too, the NFPT code conspicuously outperforms others in $ CW{\mathrm {sin}} $.

%
\begin{figure}[ht!]
\begin{center}
\includegraphics [width=8cm] {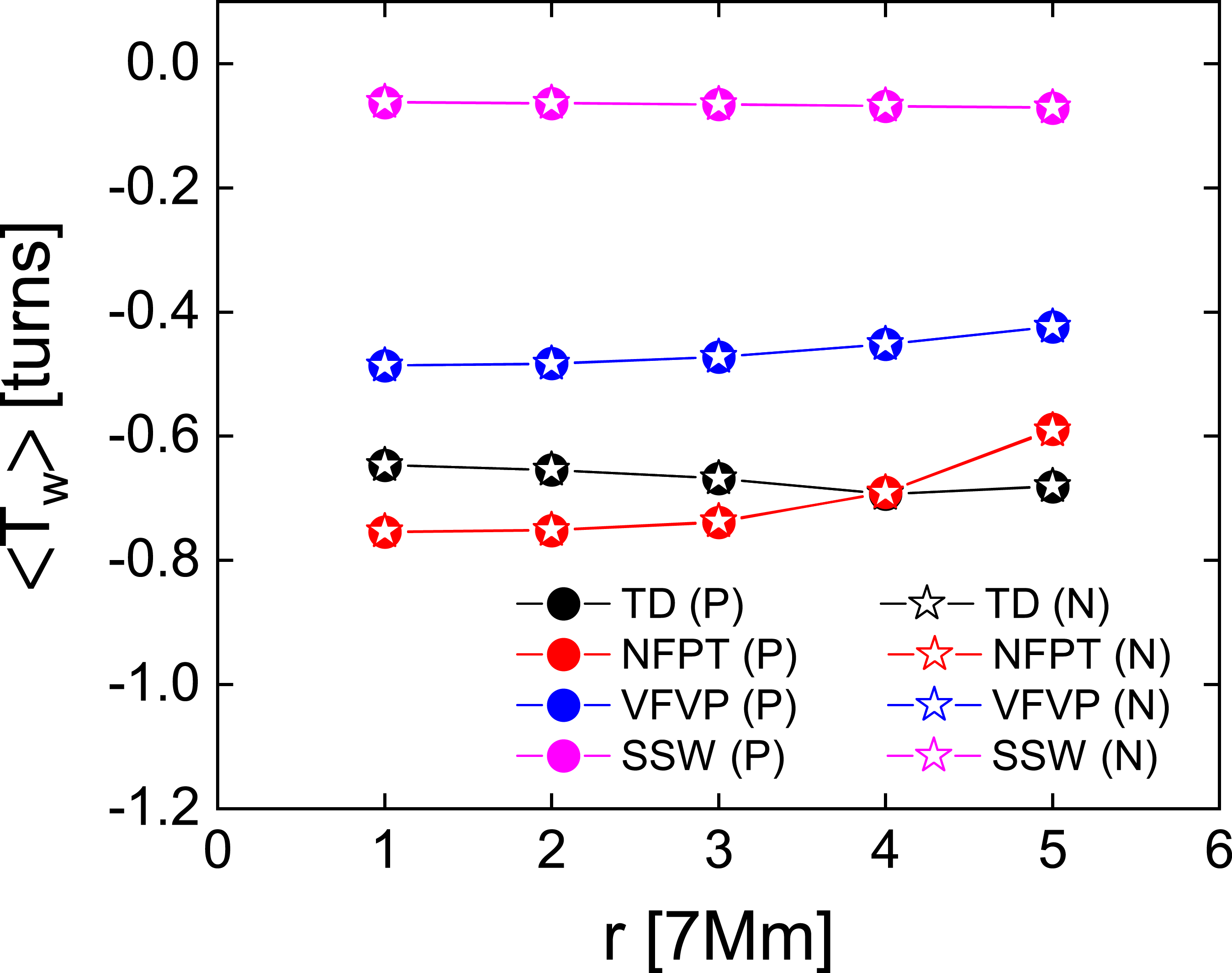}
\end{center}
\caption{Mean twists of different solutions in units of turns for TD1. 
Twenty four field line footpoints are chosen in each circle of radius $r / r_0 = 1, 2, 3, 4, 5$, where $r_0 = 7\, {\mathrm {Mm}}$, in both positive (P) and negative (N)
polarity sides, and an average is taken to give the mean twist as a function of $r$ at $z=0$. The mean twists of field lines traced from the positive footpoints and those from
the negative footpoints are almost indistinguishable. 
The twist of NFPT is quite close to that of the analytical solution (TD). The SSW solution shows the least twist in magnitude. 
\label{fig:td1twist}}
\end{figure}

Figure~\ref{fig:td1} shows that field lines are more and more loosely wound as we go from (b) to (d). 
To compare the performance of different numerical methods, we have also compared the twist of the numerical solutions with that of the analytical one. 
The twist of a field line denoted by $C_B$ is only meaningfully defined about another field line denoted by $C_A$, and 
we thus denote the twist of $C_B$ around $C_A$ by $Tw (C_A, C_B) $. In the case of a flux rope, its magnetic axis would be the most meaningful choice for $C_A$. 
In numerical solutions, the magnetic axes traced from the positive polarity and from the negative polarity can be different 
if more than one flux tube appear in a numerical solution as in the cases of VFVP and SSW. 
Thus we have to calculate two sets of twist with two magnetic axes traced from different polarity areas. 
The twist is solely defined by serial local connections between two curves, and the following formula is valid for open curves as well as for closed curves 
\citep{Tyson91, Berger06}. 
\begin{equation} \label{eq:twist}
Tw\, (C_A,C_B)=  { 1 \over {2 \pi} } \int_{C_A}  {\hat {\bm t}}_A \times {\hat {\bm u}} \cdot { {d {\hat {\bm u}}} \over {ds} }  ds\, ,
\end{equation}
in which $\displaystyle {\hat {\bm t}}_A = {  { d {\bm r}_A  } \over { ds  }  }$, where $s$ is the arclength of the curve $C_A$, is a unit tangent vector to $C_A$, and 
$\displaystyle  {\hat {\bm u}} =  {\bm u} / | {\bm u}  | $, where ${\bm u}$ is a vector connecting a point in $C_A$ to a close point in $C_B$ 
in such a way that $ {\bm u} \cdot {\hat {\bm t}}_A = 0$. 
With this method, we have measured the twist of 24 field lines, whose footpoints are in a circle of radius $r$ centered at a magnetic axis footpoint at $z=0$ , 
and have taken an average over them. 
\begin{equation}
\left< Tw(r) \right> = { { \sum\limits_{i=1}^{24}  Tw(r, \theta_i)\, \left| B_z(r, \theta_i, z=0) \right|   } \over {\sum\limits_{i=1}^{24}  \left| B_z(r, \theta_i, z=0) \right|  } }
\, ,
 \label{eq:meantwist}
\end{equation}
in which $(r, \theta_i)$ is the polar coordinates of the $i$-th field line footpoint, and $\theta_i = i \pi / 12$.  
Circles of five different radii, $r / r_0 = 1, 2, 3, 4, 5$ where $r_0 = 7\, {\mathrm {Mm}}$, are taken in the positive and negative polarity areas, respectively, 
so that ten mean values of twist in units of turns ($2\pi$ radians) are obtained for each solutions and 
are plotted in Figure~\ref{fig:td1twist}. As shown in the figure, the mean twists of field lines traced from 
the positive polarity side and from the negative side for the same $r$ are indistinguishable.  
It is remarkable that the twists of the NFPT solution 
are closest to those of the analytical solution. The twists obtained by VFVP fall a little short of the twists of the former two solutions, and 
the SSW yields quite small twists as 
seen in Figure~\ref{fig:td1}.  

%
\begin{figure*}[ht!]
\begin{center}
\includegraphics[width=18cm] {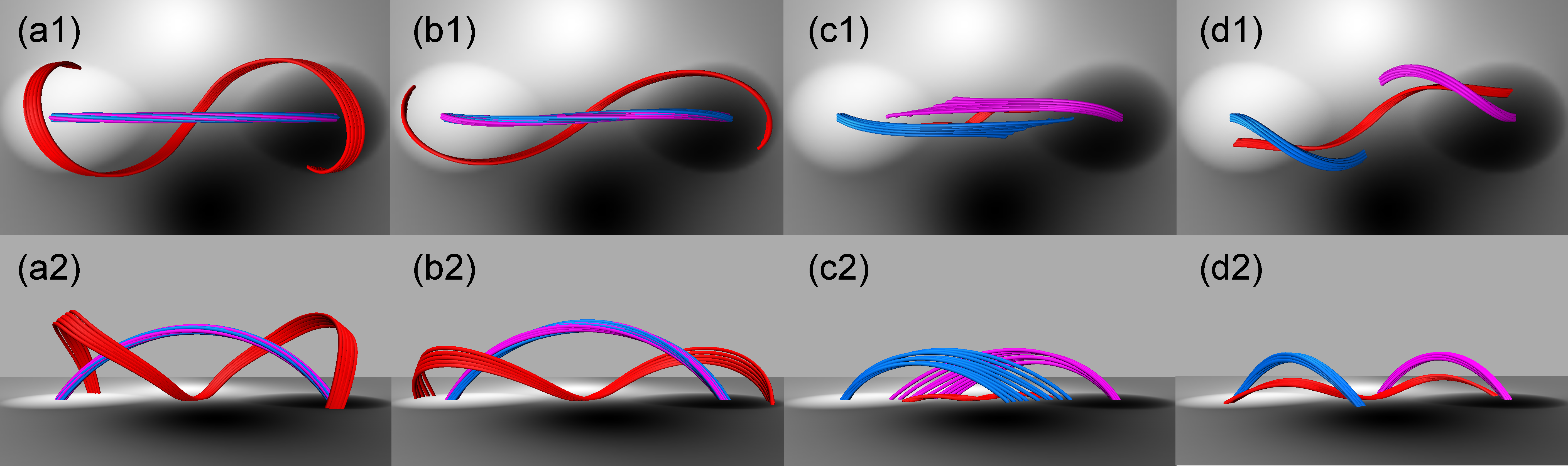}
\end {center}
\caption{Field lines of model TD2. The upper row (1) shows top views and the lower row (2) side views of the magnetic axis field lines in blue and pink  
and the bald patch field lines in red. Each column represents (a) the analytical model,  
(b) the numerical solution by our new code NFPT (non-variational FFF code in poloidal-toroidal formulation), 
(c) the numerical solution by our earlier code VFVP (variational FFF code in vector potential formulation), and 
(d) the numerical solution by the optimization code in SolarSoftWare (SSW). 
The black and white background brightness in the photosphere represents the polarity of the line-of-sight magnetic field component, i.e., white for positive and 
black for negative. 
The magnetic axis field lines in all models are traced from the same positive (blue) and 
negative (pink) footpoints. The bald patch field lines (red) are traced from the bald patch in the polarity inversion line. 
\label{fig:td2}}
\end{figure*}

Previously, the twist of a field line was sometimes measured by the following formula \citep{Inoue11, Inoue14} 
\begin{equation} \label{eq:inouetwist}
Tw  = { 1 \over {4 \pi} } \int_{C_B} \alpha \, dl
\, ,
\end{equation}
in which  $\displaystyle \alpha = {\bm B} \cdot  {\bm J}   /  B^2 $ and 
and $l$ is the arclength of the field line $C_B$. 
This method is possibly subject to criticism in that a twist cannot be defined for one curve alone. With an implicit assumption that a magnetic 
axis of a flux rope takes the role of $C_A$, the equation can be made meaningful with a correction that the line integral should be taken over the magnetic axis $C_A$, 
$dl$ being the arclength of $C_A$,  and necessarily under the condition that 
the toroidal 
current density should be uniform over each cross-sectional area of the flux tube where the field line $C_B$ lies. 
We have compared the twists by equation~(\ref{eq:twist}) and those by equation~(\ref{eq:inouetwist}) without modification for all the solutions of TD1, 
and have found that the latter method 
tends to overestimate the twists, which is attributed to the larger length of $C_B$ than $C_A$.   
The discrepancy is found to grow with $r$, because the length ratio of $C_B$ to $C_A$ 
increases with $r$. 

\begin{deluxetable*}{cccccccc}
\tablenum{3}
\tablecaption{Performance Metrics of Different Solutions for Model TD2 \label{tb:td2metric}}
\tablewidth{0pt}
\tablehead{
\colhead{Solution} &
\colhead{$ C_{\mathrm {vec}} $} &
\colhead{$ C_{\mathrm {CS}} $} &
\colhead{$ E_{\mathrm n}^{'} $} &
\colhead{$ E_{\mathrm m}^{'} $} &
\colhead{$ \epsilon $}&
\colhead{$ \epsilon_p $}&
\colhead{$ CW{\mathrm {sin}} $}
}
\startdata
Analytical  & $1.000$ & $1.000$ & $1.000$ & $1.000$ & $1.000$ & $2.125$ & $0.127$\tablenotemark{a} \\
NFPT & $0.985$ & $0.987$ & $0.832$ & $0.833$ & $0.803$ & $1.953$ & $0.045$   \\
VFVP  & $0.923$ & $0.931$ & $0.630$ & $0.627$ & $0.605$ & $1.288$ & $0.172$   \\
SSW  & $0.891$ & $0.943$ & $0.569$ & $0.560$ & $0.482$ & $1.024$ & $0.366$   \\
\enddata
\tablenotetext{a}{This value is not zero in a numerical grid. }
\end{deluxetable*}
%

%
%
\begin{figure*}[ht!]
\begin{center}
\includegraphics [width=15cm] {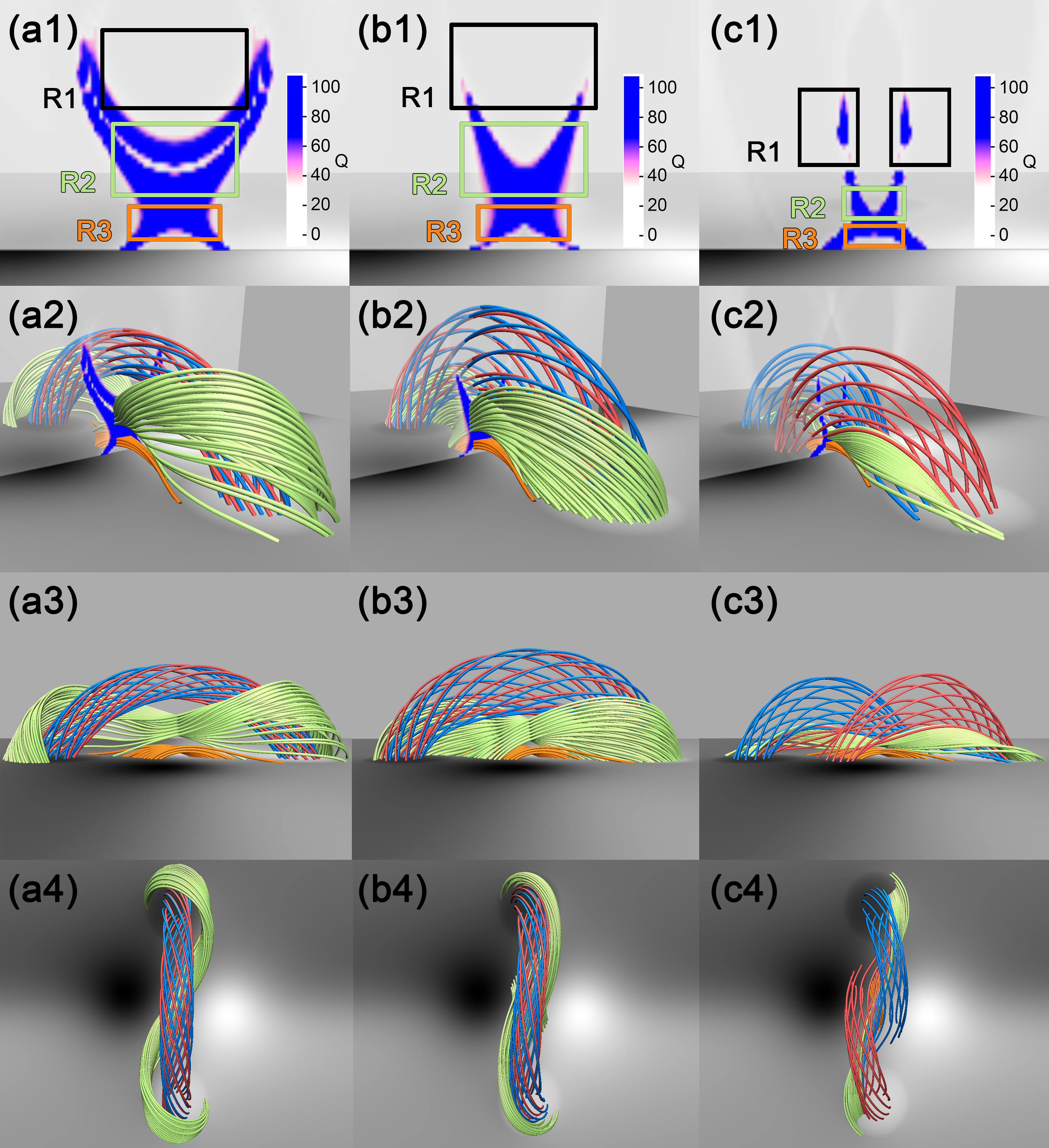}
\end{center}
\caption{The squashing factor $Q$ maps in the $y=0$ plane and field lines of TD3 showing QSLs and HFTs. Each column regards (a) the analytical solution, (b) the NFPT solution and 
(c) the SSW solution. The first row shows $Q$ color maps and three boxed areas, through which the field lines shown in other rows are passing. 
The black box (R1) lies above the quasi-separator (QS: magnetic axis of the HFT), the light green box (R2) encloses the QS and the orange box (R3) lies below the QS. 
The second to fourth rows show three sets of field lines passing through the three boxes in different perspectives. 
The field lines through R1 represent the TD flux rope. They are given in red when field lines are traced from the positive polarity side of the flux rope and in blue when traced from the negative polarity side. The analytical and NFPT solutions show one flux rope, but the SSW solution shows two flux ropes.
The field lines through R2 given in light green form a part of the QSL. The QSL structure is most conspicuously shown in (a), then in (b) and faintly in (c). 
\label{fig:td3}}
\end{figure*}

The model TD2 (see Table~\ref{tb:tdpara}) characteristically has a bald patch, which is a segment in a polarity inversion line, where locally concave upward field lines touch 
the photosphere \citep{Titov93}. A bald patch is a possible location of a solar prominence of inverse polarity \citep{Lee95, Mackay10}, and it can develop into a current sheet 
\citep{Low92, Cheng98}, where magnetic reconnection may take place \citep{Delannee99}. 
In TD models, a bald patch appears only with a sufficiently large twist of the flux rope \citep{Titov99}. TD models with bald patches have been 
reproduced in several numerical solutions \citep{Valori10, Jiang16, Demcsak20}. 

The parameters of our TD2 are the same as those of Case~5 in \citet{Demcsak20}, in which a Grad-Rubin type code \citep{Wheatland07} was used. Figure~\ref{fig:td2} shows the top and side views of four TD2 solutions, (a) analytical, (b) NFPT, (c) VFVP and (d) SSW. As in the case of TD1, the NFPT solution reproduces one flux rope with a single magnetic axis of the analytical solution. The solutions by VFVP and SSW produce two flux tubes. The two flux tubes by SSW are widely separated. All four models show bald patches, but the endpoints of the bald patch field lines are quite differently located. The footpoint positions of the bald patch field lines apparently indicate their writhes \citep{Berger06}. The footpoints and the S-shape appearances of the bald patch field lines in the NFPT solution and the analytical solution are quite close although the field lines in the latter are slightly longer and their apex reaches a higher altitude. While the separation of the magnetic axes of two flux tubes is smaller in VFVP than in SSW, the writhe of the bald patch field lines is larger in SSW than in VFVP. Compared with the solution by \citet{Demcsak20}, the analytical solution and and the NFPT solution show 
larger writhes of bald patch field lines than their result while the VFVP and SSW solutions show smaller writhes. Limited to TD models, nonvariational methods (NVPT and a Grad-Rubin type code) seem to be better at revealing topological features of the field than variational methods (VFVP and SSW). 

The performance metrics of the four solutions for TD2 are listed in Table~\ref{tb:td2metric}. 
The NFPT solution exceeds other numerical solutions in all metric scores, and it particularly stands out in $ CW{\mathrm {sin}} $. 
It is again suggested that the force-freeness represented by $ CW{\mathrm {sin}} $ is more important in reproducing 
topological features of the fields, e.g., flux ropes and bald patches, than other metrics.  

%


\subsection{Hyperbolic Flux Tubes in the Titov-D\'{e}moulin Models} \label{subsec:tdhft}

If a magnetic field is purely two-dimensional, i.e., if it depends on two coordinates and its field lines lie in parallel planes, an X-shaped field configuration 
has two surfaces called separatrices intersecting each other at a line called a separator. 
The field lines in each quadrant may locally be approximated by hyperbolas near the X-point and they have discontinuous connectivities across separatrices. The separator may be 
deformed
into a current sheet, where magnetic reconnection can take place
\newpage
\begin{deluxetable*}{cccccccccc}
\tablenum{4}
\tablecaption{Performance Metrics of Different Solutions for Model TD3 \label{tb:td3metric}}
\tablewidth{0pt}
\tablehead{
\colhead{Solution} &
\colhead{$ C_{\mathrm {vec} } $} &
\colhead{$ C_{\mathrm {CS} } $} &
\colhead{$ E_{\mathrm n}^{'} $} &
\colhead{$ E_{\mathrm m}^{'} $} &
\colhead{$ \epsilon $}&
\colhead{$ \epsilon_p $}&
\colhead{$ CW{\mathrm {sin} } $} &
\colhead{$ E_{q, {\mathrm {bot}}}$\tablenotemark{a}} &
\colhead{$E_{q, {\mathrm {whole}}}$\tablenotemark{b}} 
}
\startdata
Analytical  & $1.000$ & $1.000$ & $1.000$ & $1.000$ & $1.000$ & $2.542$ & $0.121$ & $0$ & $0$  \\
NFPT & $0.978$ & $0.968$ & $0.765$ & $0.725$ & $0.788$ & $2.009$ & $0.094$ & $0.670$  & $0.813$ \\
SSW  & $0.958$ & $0.972$ & $0.672$ & $0.586$ & $0.659$ & $1.675$ & $0.388$  & $3.736$ & $1.357$ 
\enddata
\tablenotetext{a}{The error metric for the $Q$ values at the bottom boundary only.}
\tablenotetext{b}{The error metric for the $Q$ values in the whole computational domain.}
\end{deluxetable*}
%
%
\begin{figure*}[ht!]
\begin{center}
\includegraphics[width=15cm] {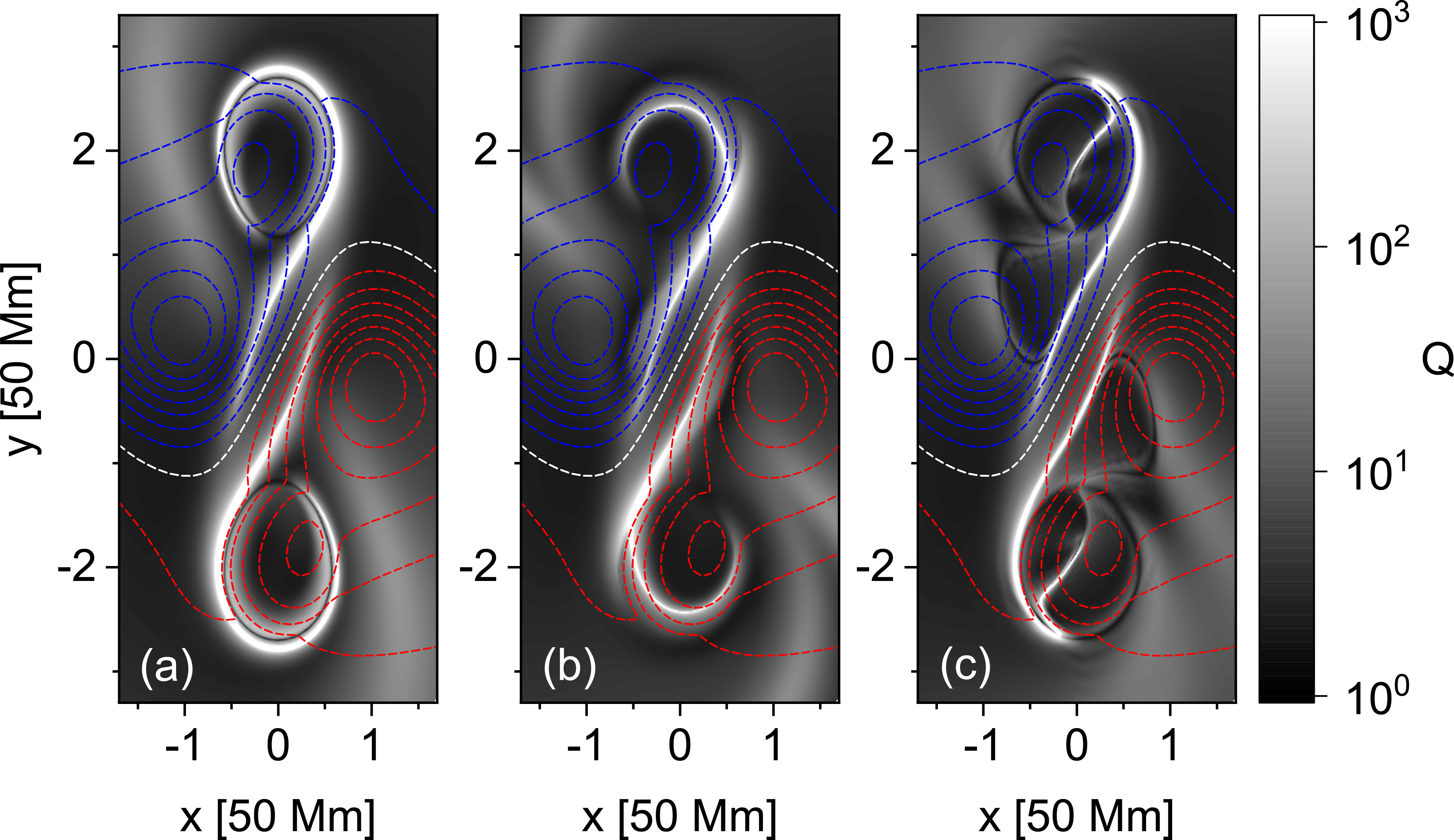}
\end{center}
\caption{The squashing factor $Q$ maps in the $z=0$ plane in black and white brightness and the isocontours of $B_z$ for (a) the analytical solution, (b) the NFPT solution and 
(c) the SSW solution for TD3. The red contours represent $B_z>0$ and the blue ones $B_z < 0$. 
\label{fig:td3qbot}}
\end{figure*}
%
\begin{figure}[ht!]
\begin{center}
\includegraphics[width=8cm] {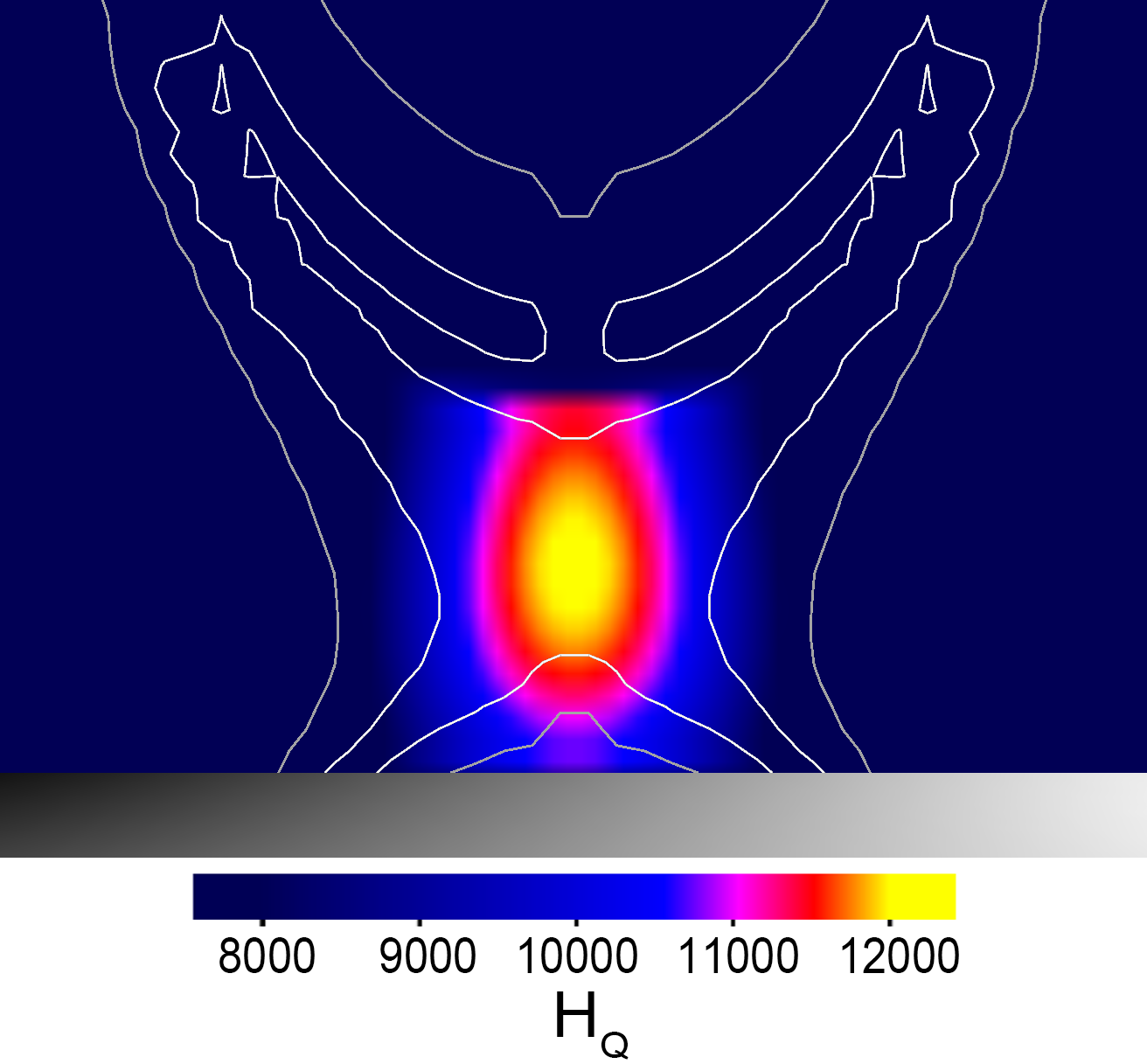}
\caption{The color map of $H_Q$ in the $y=0$ plane superposed with the isocontours of $Q$ for the analytical solution of TD3. While the $Q$ contours delineate the cross-sectional structure of a QSL, the $H_Q$ map quite well reveals the quasi-separator, which is the magnetic axis of an HFT. \label{fig:hqy0}}
\end{center}
\end{figure} 
\begin{figure*}[ht!]
\begin{center}
\includegraphics[width=15cm] {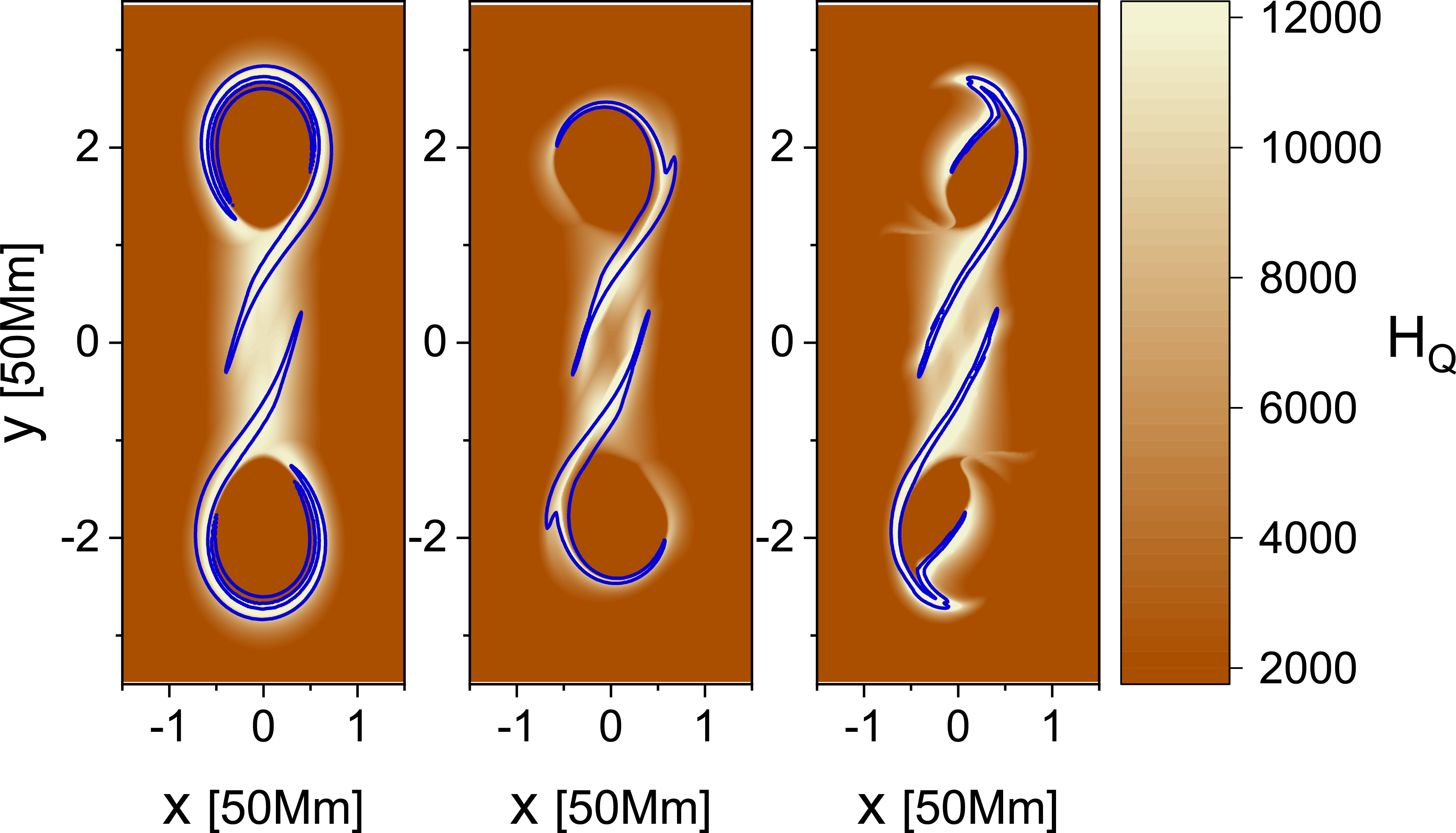}
\end{center}
\caption{The maximum $H_Q$ value of each field line rooted in the $z=0$ plane given in brownish brightness and the blue contours for $Q=100$ for (a) the analytical solution, 
(b) the NFPT solution and (c) the SSW solution for TD3. The $H_Q$ maps and the $Q$ contours show common features. 
\label{fig:hqz0}}
\end{figure*}
\noindent
\citep{Syrovatskii81}. 
If we add a smooth third component of magnetic field dependent on two coordinates to a purely 2D magnetic field, two field lines, whose endpoints are very close with a former 
separatrix between them, have a large separation at the other ends \citep{Longcope94}. Thus, the field connectivity is everywhere continuous. 
In a 3D situation, separatrices and separators appear if nullpoints or bald patches are present in the domain. Otherwise, the field connectivity is everywhere continuous, but 
there may exist quasi-separatrix layers (QSLs), across which the separation between the ends of field lines, which are very close somewhere in the domain, is very large 
\citep{Priest92, Demoulin96}. 
Quite similarly to the 2.5D situation, the field lines in the vicinity of the intersection of two QSLs look like hyperbolas when they are projected onto the plane normal to the intersection. 
This structure is called a hyperbolic flux tubes (HFTs) and its magnetic axis is a quasi-separator \citep{Titov02, Titov03}. A current sheet is likely to be formed in an HFT by magnetic pinching induced 
by suitable footpoint motions and consequently magnetic reconnection ensues \citep{Titov03}. To find an HFT, \citet{Titov02} proposed a squashing factor (or degree) $Q$, which represents 
how much a cross-section of a flux tube is deformed at the other side, as follows. 
\begin{equation}
\label{eq:qfactor}
Q = \dfrac 
{  \left( {\partial X_{\mp} } \over {\partial x_{ \pm}  } \right)^2   
+ \left( {\partial X_{\mp} } \over {\partial y_{ \pm}  } \right)^2   
+ \left( {\partial Y_{\mp} } \over {\partial x_{ \pm}  } \right)^2 
+ \left( {\partial Y_{\mp} } \over {\partial y_{ \pm}  } \right)^2     
}
{     \left|  \left( {\partial X_{\mp} } \over {\partial x_{ \pm}  } \right)
               \left( {\partial Y_{\mp} } \over {\partial y_{ \pm}  } \right)
          -   \left( {\partial X_{\mp} } \over {\partial y_{ \pm}  } \right)
              \left( {\partial Y_{\mp} } \over {\partial x_{ \pm}  } \right)
     \right| 
} \, ,
\end{equation}
in which $\left[ X_{\mp} (x_{ \pm}, y_{ \pm}),  Y_{\mp} (x_{ \pm}, y_{ \pm})  \right] $ is a vector function connecting a footpoint 
$(x_{ \pm}, y_{ \pm})$ to its conjugate footpoint $(x_{ \mp}, y_{ \mp})$. The flux surface of an HFT is an isosurface of $Q$ with $Q \gg 2$. 

To see how well numerical FFF solvers reproduce a TD field containing an HFT, we have constructed the model TD3 in Table~\ref{tb:tdpara} and 
calculated the squashing factor $Q$ 
by tracing field lines to the boundary from every grid-point in the domain. 
The upper row of Figure~\ref{fig:td3} shows the $Q$ value in the $y=0$ plane, which is a vertical plane cutting the TD flux rope at its apex. Apparently the $Q$ color maps show
an X-shaped configuration, which is a typical feature of an HFT. The X-shaped structure manifested by the same color scheme of $Q$ is 
most conspicuous in the analytical solution, then in the NFPT solution and rather faint 
in the SSW solution. Figure~\ref{fig:td3} also shows field lines passing through three boxed areas in the $y=0$ plane, a black box above the axis of HFT, a light green box in the 
vicinity of the HFT axis and an orange box just below it. The upper field lines above the HFT form one flux rope in the analytical and NFPT solutions and two flux ropes in the SSW
solution. The light green field lines are supposed to cover a part of the QSL. The structure of the QSL is best revealed in the analytical solution and then in the NFPT solution. 

Figure~\ref{fig:td3qbot} shows the $Q$ value in the $z=0$ plane in black and white brightness, superposed with the $B_z$ level contours, positive in red and negative in blue. 
In all three solutions, the $Q$ distributions below the central part of the flux ropes ($ -1 < y <1 $) are quite similar, but in the vicinity of the flux rope legs, the $Q$ distributions are 
noticeably different. The $Q$ map of the analytical solution has hook-like structures, which almost surround the flux rope footpoint areas. The result of our NFPT
also shows hook-like structures, which, however, are a little smaller and thinner than the analytical ones. In the $Q$ map by SSW, the hooks are not fully wound, but are broken in the
middle.  
Apparently the difference in the $Q$ maps of TD3 is somewhat similar to the difference in the bald patch field lines of TD2 shown in Figure~\ref{fig:td2}. 
To assess the proximity of the overall $Q$ distribution of a numerical model to the analytical one, we devise the following metric
\begin{equation} \label{eq:eq}
E_q=\displaystyle{{{1\over M} \sum\limits_{i}{{\left|q_{i}-Q_{i}\right|}\over{\left|q_{i}\right|}}}}\, ,
\end{equation}
where $i$ is the grid-point index, $M$ the total number of grid-points, $q$ the squashing factor of the analytic model, and $Q$ the squashing factor of a numerical solution.
If the numerical solution is identical to the analytical one, the metric $E_q$ should be zero. We have evaluated the values of $E_q$ in the $z=0$ plane ($E_{q, {\mathrm {bot}}}$) 
as well as in the whole computational domain ($E_{q, {\mathrm {whole}}}$) and have listed them in Table~\ref{tb:td3metric} along with the performance metrics. 
The error metric $E_{q, {\mathrm {bot}}}$ of NFPT is about one fifth that of SSW. However, the error metric $E_{q, {\mathrm {whole}}}$ does not show so 
much difference as $E_{q, {\mathrm {bot}}}$, which may be attributed to the larger portion of grid-points with small $q$ ($Q$ also) in the whole domain 
than in the bottom plane only.

In an HFT, the $Q$ value takes a maximum in a line called a quasi-separator \citep{Titov02}. It corresponds to an X-line in a 2.5D magnetic field and may be called the magnetic
axis of an HFT. If one purposes to find a quasi-separator (QS) and the HFT in its neighborhood rather than to find the entire structure of a QSL, one may not want to take the trouble of
evaluating $Q$, which requires finding the conjugate footpoint pairs by field line tracing. Here we propose a rather simple ``local'' method of locating a QS and HFT. 
In a plane normal to a QS, field lines projected onto this plane are like hyperbolas and the magnetic field in this plane has the following property, 
\begin{equation} \label{eq:jacobp}
{\mathcal J}_{\perp} = 
\renewcommand\arraystretch{1.7}
\begin{vmatrix}
\displaystyle{{\partial{B_{x'}}} \over {\partial{x'}}} &\displaystyle{{\partial{B_{x'}}} \over {\partial{y'}}}\\
\displaystyle{{\partial{B_{y'}}} \over {\partial{x'}}} &\displaystyle{{\partial{B_{y'}}} \over {\partial{y'}}}
\end{vmatrix}
<0 \, ,
\end{equation}
in which $x'$ and $y'$ are arbitrary Cartesian coordinates in this plane. 
Since we do not know the location of the QS from the beginning, we just calculate ${\mathcal J}_\perp$ in a plane normal to the local magnetic field. Then, we set 
\begin{equation} \label{eq:hq}
H_Q = 
\begin{cases}
  -{\mathcal J}_\perp  & \text{if} \ {\mathcal J}_\perp < 0 \\
  0              & \text{if} \  {\mathcal J}_\perp \ge 0
\end{cases}
\end{equation}
so that we may focus on candidate points for a QS. 
Although the isocontours of $H_Q$ would not follow the shape of a QSL, the maximum of $H_Q$ is expected to be located on the QS. 
Figure~\ref{fig:hqy0} shows a color map of $H_Q$ in the $y=0$ plane made from the analytical solution of TD3 superposed with white isocontours of $Q$. 
While $Q$ contours are useful for identifying a QSL structure, an $H_Q$ map (or contours) is advantageous for locating a QS. 
As both $Q$ and $H_Q$ are useful for finding an HFT, their distributions in the bottom plane ($z=0$) are expected to be similar. This is confirmed 
by Figure~\ref{fig:hqz0}, in which the maximum value of $H_Q$ in a field line is expressed by brightness of brown color at its footpoint in $z=0$ and the isocontours of 
$Q=100$ are given in blue for the analytical solution, the NFPT solution and the SSW solution of TD3. 
The figure shows that the distributions of the two quantities are quite consistent. For practical purposes, we do not need to find the maximum of $H_Q$ is each field line. 
One may roughly draw field lines globally and construct an $H_Q$ distribution in a suspected region. 
Then, $H_Q$ can be a quick tool for locating a QS and an HFT. 
%


\section{Discussion and Summary} 
\label{sec:sum}

\subsection{Efficiency and Effectiveness of the Computational Code} \label{subsec:effcode}

Our NFPT code requires solving two 3D Poisson equations (\ref{eq:bziter}) and (\ref{eq:jziter}) and two 2D Poisson equations (\ref{eq:phiupd}) and (\ref{eq:azupd}) 
for $k=1,2, ..., N_z$ planes in each iteration step. In the code, the Poisson equations are solved by a direct solver package FISHPACK \citep{Swarztrauber75} 
quite efficiently and accurately. The computational time per iteration step required by NFPT is about ten times that of VFVP and about twice that of SSW. 
However, the number of iteration steps required for convergence by NFPT is about one tenth of that by VFVP or SSW. In our TD2 calculations, the NFPT code required about 200 
iteration steps while the other codes demanded at least 2500 steps. In terms of the computational resources required for producing a solution, the NFPT is 
comparable to the VFVP and uses far less resources than the SSW. 

A problem described by a set of partial differential equations is well-posed if the imposed boundary conditions and constraints yield a unique solution. 
Although a problem is mathematically well-posed, its numerical implementation may not be so for diverse reasons. Among others is the 
way of information transfer between the boundary and the inner computational domain, inherent to the numerical method.  
In order to see whether the boundary condition is well reflected in the solution consistently throughout the computational domain, we have devised the following test. 
Suppose that a numerical solution has been constructed in a domain
\begin{align*}
 V_0 = \{ (x,y,z) \, | \, & {0 \le x \le L_x ,} \, {0  \le y \le L_y,} \\
                                   & {0 \le z \le L_z} \}
\end{align*}
with a numerical grid 
\begin{align*}
 G_0 = \{ (i,j,k) \, | \, & i = 0, 1, 2, \ldots, N_x ; \  j = 0, 1, 2, \ldots, N_y  ; \\
                                & k = 0, 1, 2, \ldots, N_z \} \, .
\end{align*}
Then, one can construct a numerical solution 
in a domain reduced in $z$, i.e., in 
\begin{align*}
V_{z_0} = \{ (x,y,z) | \, & { 0 \le x \le L_x ,} \, {0  \le y \le L_y ,} \\
                                  & {0 < z_0 \le z \le L_z} \} 
\end{align*}
with a numerical grid
\begin{align*}
G_{k_0} = \{ (i,j,k) \, | \, & i = 0, 1, 2, \ldots, N_x ; \  j = 0, 1, 2, \ldots, N_y ;  \\ 
                                   & k = k_0, k_0+1, k_0+2,\ldots, N_z \},  
\end{align*}
where $k_0 \Delta z = z_0$ with $\Delta z = L_z / N_z$,  
using the former numerical solution at  
$z= z_0$ ($k=k_0$) as a boundary condition at the bottom boundary of the new domain. At this time, we should not use the former solution as the initial condition, 
nor impose any known solutions as 
the boundary conditions at the lateral and top boundaries of $G_{k_0}$. If the numerical solutions with grids $G_0$ and $G_{k_0}$ are identical in the domain $G_{k_0}$, it can 
be said that the bottom boundary and the solution in the domain are consistently connected. 
For TD2, we have employed a numerical grid with $(N_x, N_y, N_z) = (150, 250, 100)$. To perform the above test, we have chosen $k_0=10, 20, 30, 40, 50$ and 
constructed numerical solutions with the NFPT code in five different grids $G_{k_0}$, in all of which the numerical solutions obtained in the original grid $G_0$ are used
as bottom boundary conditions at $k=k_0$ of $G_{k_0}$.  
To compare the numerical solutions in the original grid and in each reduced grid, we have evaluated five performance metrics by \citet{Schrijver06} in the domain $G_{k_0}$,  
using the solution 
obtained in each reduced grid $G_{k_0}$ for ${\bm B}$ and the solution in the original grid $G_0$ for ${\bm b}$ 
in equations~(\ref{eq:cvec})--(\ref{eq:eps}). The results are given in Table~\ref{tb:reduced}. 
The metrics $C_{\mathrm  {vec}}$ and $C_{\mathrm {CS}}$ are $0.999$ for all $k_0$ while other metrics 
show a slight tendency of degradation with increasing $k_0$. These scores of NFPT are unrivaled with those of variational codes (VFVP and SSW), all the more so with increasing $k_0$, unless 
fixed values of the field are prescribed at all boundaries. 
Thus, it can be said that the PT representation and its natural way of imposing boundary conditions are self-consistent to
generate an unambiguous solution in the whole domain. It is also suggested that this test be performed on other numerical methods to be developed. 

\begin{deluxetable}{cccccc}
\tablenum{5}
\tablecaption{Performance Metrics of the NFPT Solutions for Model TD2 in Reduced Grids \label{tb:reduced}}
\tablewidth{0pt}
\tablehead{
\colhead{$k_0$} &
\colhead{$ C_{\mathrm {vec}} $} &
\colhead{$ C_{\mathrm {CS}} $} &
\colhead{$ E_{\mathrm n}^{'} $} &
\colhead{$ E_{\mathrm m}^{'} $} &
\colhead{$ \epsilon $}
}
\startdata
 $10$ & $0.999$ & $0.999$ & $0.955$ & $0.953$ & $0.948$  \\
 $20$ & $0.999$ & $0.999$ & $0.955$ & $0.949$ & $0.959$    \\
 $30$ & $0.999$ & $0.999$ & $0.947$ & $0.942$ & $0.944$    \\
 $40$ & $0.999$ & $0.999$ & $0.936$ & $0.931$ & $0.914$   \\
 $50$ & $0.999$ & $0.999$ & $0.924$ & $0.919$ & $0.890$   
\enddata
\tablecomments{For all the metrics, the numerical solution in the grid $G_0$ is taken for the reference solution ${\bm b}$ and the numerical solution in the grid $G_{k_0}$ 
is taken for ${\bm B}$. }
\end{deluxetable}

\subsection{Discussion on Lateral and Top Boundary Conditions} \label{subsec:dissbc}

\begin{deluxetable*}{cccccccccccc}
\tablenum{6}
\tablecaption{Boundary Fluxes and Performance Metrics for Different Boundary Conditions \label{tb:bccase}}
\tablewidth{0pt}
\tablehead{
\colhead{BC} &
\colhead{$F_l / F_b $} &
\colhead{$F_t / F_b $} &
\colhead{$(F_l + F_t) / F_b $} &
\colhead{$ C_{\mathrm {vec} } $} &
\colhead{$ C_{\mathrm {CS} } $} &
\colhead{$ E_{\mathrm n}^{'} $} &
\colhead{$ E_{\mathrm m}^{'} $} &
\colhead{$ CW{\mathrm {sin} } $} &
\colhead{$[F_l / F_b]_d $} &
\colhead{$[F_t / F_b]_d $} &
\colhead{$[(F_l + F_t) / F_b]_d $} 
}
\startdata
BC1  & $0.0$   & $0.259$  & $0.259$ & $0.985$ & $0.987$ & $0.832$ & $0.833$ & $0.0450$ & $0.0$   & $0.0746$  & $0.0746$ \\
BC2  & $0.244$ & $0.0534$ & $0.298$ & $0.977$ & $0.977$ & $0.807$ & $0.805$ & $0.0552$ & $0.324$ & $0.00213$ &  $0.327$ \\
BC3  & $0.424$ & $0.0630$ & $0.487$ & $0.914$ & $0.916$ & $0.617$ & $0.607$ & $0.0590$ & $0.462$ & $0.00839$ &  $0.470$ \\
BC4  & $0.329$ & $0.257$  & $0.586$ & $0.963$ & $0.960$ & $0.749$ & $0.738$ & $0.0980$  & $0.327$ & $0.0417$  &  $0.368$ 
\enddata
\tablecomments{The total unsigned magnetic flux through the bottom boundary is denoted by $F_b$, the total unsigned flux through the lateral boundary by $F_l$, and
the total unsigned flux through the top boundary by $F_t$.  The total open flux is $F_l + F_t$. The three flux ratios obtained in a domain 
with a double size in $z$ are designated by the subscript $d$.}
\end{deluxetable*}

The lateral boundary condition we have used in this paper does not allow any magnetic flux to cross the lateral boundary. Any imbalanced flux at the photospheric boundary
is thus ducted through the top boundary. Previously, other boundary conditions have been proposed, which allow nonzero normal magnetic field 
at the lateral boundary \citep[e.g.,][]{Seehafer78, Otto07}. These boundary conditions assume image polarities across the lateral boundary by a plane symmetry or a line
symmetry of a certain parity so that there may be field lines connecting the polarity patches in the real domain and those in the image domains. Thus, 
the field configuration in the real domain is more or less affected by the field connection across the lateral boundary. 
Since there are several methods of placing image polarities, the resulting field configuration depends on the method chosen \citep{Otto07}. 
By mathematical reasoning alone, one can hardly tell which boundary condition is superior to others . 
Our non-penetrating wall condition, which is also an artificial boundary condition, is not an exception. 
As far as a more or less isolated active region is concerned, however, the connections between the real (observed) polarities and the image polarities 
may rather be considered undesirable. For TD2, we have tested four different lateral boundary conditions, one of which is our non-penetrating condition, 
in order to see how much magnetic flux passes through the lateral and top boundaries and how close the numerical solution is to the reference field. 
The four lateral boundary conditions are as follows. 
\begin{enumerate}[label= (BC\arabic*) ]
\item $\displaystyle {\partial {B_z}} / {\partial n} = 0$, $\displaystyle J_z = 0$.    
\item $\displaystyle B_z = 0$, $\displaystyle {\partial {J_z}} / {\partial n} = 0$. 
\item $\displaystyle B_z = 0$, $J_z=0$. 
\item  $\displaystyle {\partial {B_z}} / {\partial n} = 0$, $\displaystyle {\partial {J_z}} / {\partial n} = 0$. 
\end{enumerate}
In BC1, which is taken in the present paper, $B_n=0$, and $J_n$ is not necessarily zero, but must be zero in a force-free state. 
In all the other cases, $B_n \ne 0$. In BC2, ${\bm B}_t = 0$, $J_n = 0$, and ${\bm J}_t$ is not necessarily zero, but must be zero in a force-free state. 
Thus, the magnetic field passing through the lateral boundary in BC2 is a potential field. For BC1 and BC2 both, the force-free coefficient $\alpha$ is anti-symmetric across 
the lateral boundary, and neither condition can be accommodated in LFFF models. Since there is no current flowing through the lateral and top boundaries with BC1 and BC2, 
the current at the bottom boundary must be preprocessed  so that the net current may be zero. In BC3 and BC4, there are normal magnetic currents as well as normal
magnetic fields crossing the lateral boundary. Since the force-free coefficient $\alpha$ is symmetric across the lateral boundary, these boundary conditions can be 
used for LFFF model as BC3 was adopted by \citet{Seehafer78}. The line symmetry boundary condition by \citet{Otto07} has not been tried here.

In Table~\ref{tb:bccase}, the total unsigned magnetic flux through the lateral boundary $F_l$, 
the total unsigned flux through the top boundary $F_t$, the total unsigned open flux 
$\displaystyle   F_{\mathrm{open}} =     F_l + F_t  $ are given in units of the total unsigned flux through the bottom boundary $F_b$.  
Also, performance metrics are given for each case. 
The total unsigned open flux is the smallest with BC1, and the lateral flux of other cases is almost comparable to that. 
The performance metrics are best with BC1, which implies that the numerical solution is closest to the reference field among other cases. 
Since the lateral magnetic flux is also expected to depend on the domain's aspect ratio, 
we have also tried to run the code in a domain, whose height is 
twice the original $L_z$ while the $x$- and $y$-sizes are unchanged. The ratios of boundary fluxes to the bottom flux in this case are given 
in the last three columns of Table~\ref{tb:bccase}. The flux through the top boundary is significantly reduced for all cases 
compared with the value in the original domain. However, the total open flux through the lateral and
top boundaries is either slightly decreased (BC3 and BC4) or even increased a little (BC2). With BC1, the total open flux is significantly decreased by doubling 
the domain size. Although not given in Table~\ref{tb:bccase}, the performance metrics are best with BC1 in the double-sized domain too. 
Based on this investigation about the boundary conditions and the code performance, it is carefully suggested that 
our boundary condition (BC1) would be the best choice for reconstruction of a rather isolated active region and that it would relieve one from worrying about where and how to 
put image polarities and current sources outside the domain.

\begin{figure}[ht!]
\begin{center}
\includegraphics[width=8cm] {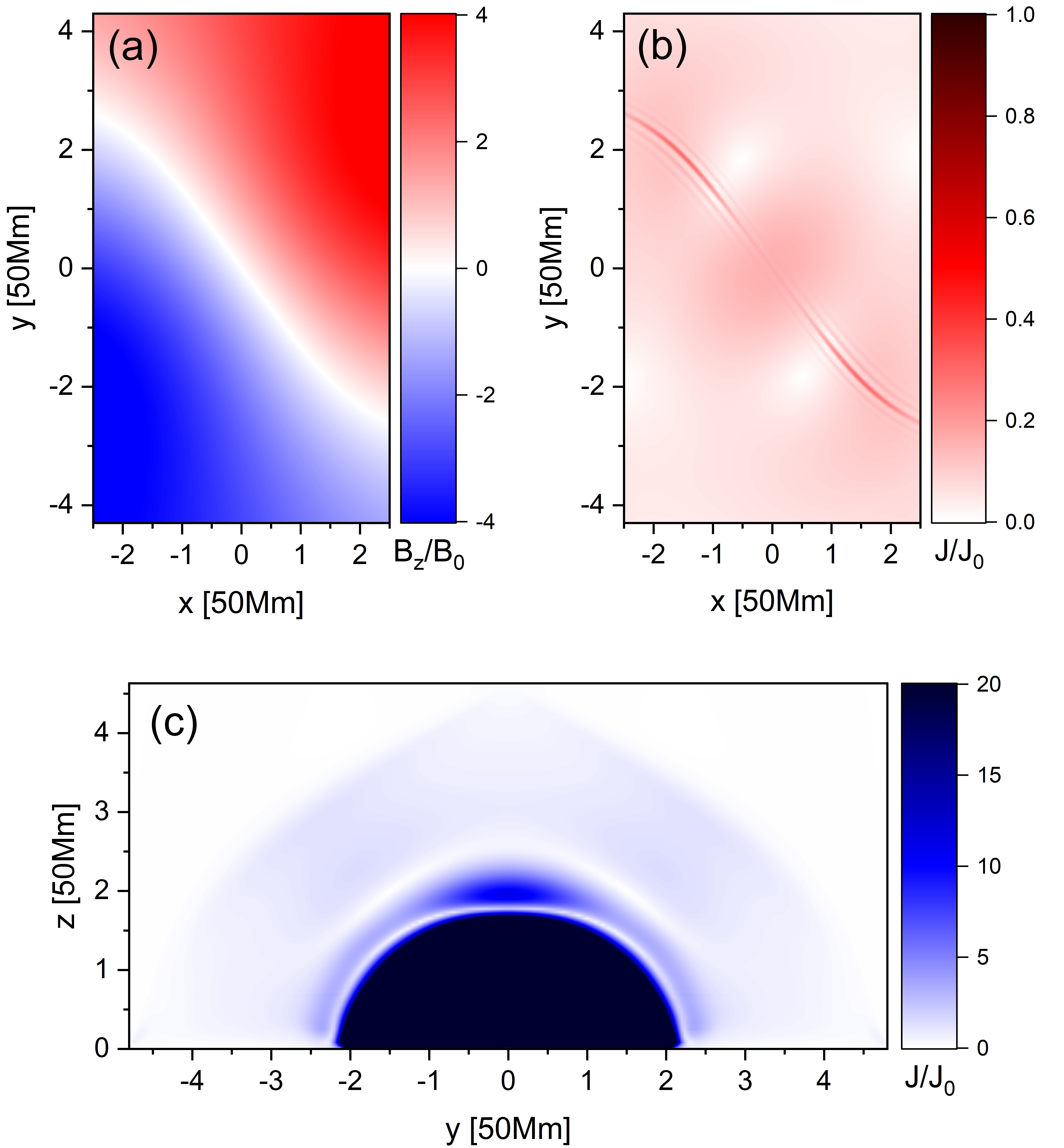}
\caption{The magnetic field null-line at the top boundary and the separatrix surface emanating from it. 
(a) The distribution of $B_z (z = L_z)$  is given as a color map. The positive polarity is given in red and the negative in blue. The null-line is given in white. 
(b) The magnitude of current density $J = | {\bm J} |$ at the top boundary is given in darkness of red. The high $J$ region delineates the null-line in (a). 
(c) The magnitude of current density $J = | {\bm J} |$ in the vertical plane $x=0$ is given in darkness of blue. The outer boundary of the blue region has a slight 
enhancement in $J$ and it indicates the intersection of the separatrix current sheet with the plane $x=0$.  \label{fig:bzj}}
\end{center}
\end{figure} 

It should be noted that the TD models have a perfect flux balance at the bottom boundary and the open flux is caused by the 
boundary conditions, not by a flux imbalance. 
When we use a source surface condition given by equations~(\ref{eq:bxbyss}) and (\ref{eq:bzss})
at the top boundary, there must be a polarity inversion line where $B_z = 0$ at this boundary unless the bottom magnetic field is totally uni-directional. 
From the source surface condition and the force-free condition, ${\bm J} (z=L_z)=0$. Thus, the open field is a potential field. A surface emanating from the null-line to the bottom boundary
is a separatrix surface, across which field topology discontinuously changes from ``closed'' to ``open'' as well as ``non-potential'' to ``potential.'' 
Since there is no reason for the field direction to be continuous across this
separatrix, this surface must be a current sheet \citep{Zwingmann85, Platt94}. As we do not know in advance where the footpoints of the separatrix are located at the bottom boundary, the footpoint line
may lie where $J_z \ne 0$ unless we have a wide enough buffer area with $J_z=0$ and $B_z \ne 0$ at the bottom boundary. This is an intrinsic problem of our model, which 
causes those performance metrics evaluating the proximity of the numerical solution to the reference field to deviate from $1$ 
even though the solution in the closed flux region is quite similar to the reference field. The metric $CW_{\mathrm {sin}}$ measuring the force-freeness, 
however, is minimally affected 
except near the separatrix. To see whether the null-line with a singular transverse current exists at the top boundary and whether a separatrix current sheet exists in the domain, 
we have plotted the distributions of $B_z$ and $J = | {\bm J} |$ at the top boundary and that of $J$ in the vertical plane $x=0$ in Figure~\ref{fig:bzj}. 
Because the value of numerically evaluated current density highly depends on magnetic field strength and the null-line and the separatrix are in a weak field region, 
the features we are seeking 
do not stand out conspicuously, but they are still identifiable in the figure. 
As we have expected, the major field structure with strong currents lies in the closed flux region, 
and we have thus been able to reproduce the significant geometrical features of the reference TD fields as given in the previous section.

\subsection{Summary} \label{subsec:summary}

In summary, we have presented  a new method of constructing a coronal force-free field based on a poloidal-toroidal representation of magnetic field. The PT representation 
allows us to impose the boundary conditions $B_n$ and $J_n$ at the photospheric boundary once and for all with only the boundary values of the poloidal and toroidal functions. 
With a rigid, conducting, slip wall boundary conditions at the lateral boundaries and a source surface condition at the top boundary, a magnetic flux imbalance at 
the bottom boundary can be accommodated. Since no current can escape the computational domain, the current at the bottom boundary must be preprocessed so that the 
net current through it may be zero. At the top boundary, a rigid, conducting, slip wall condition can also be used instead of the source surface condition. With this condition, however, 
a flux imbalance at the bottom boundary is not allowed and thus the $B_n$ data there must be preprocessed. Our new method is nonvariational in the sense that the converging 
sequence toward the solution does not extremize any conceivable functional. It rather directly targets a solution, but iterations are needed because of the high degree of nonlinearity. 
Thus, it requires far fewer iteration steps than variational methods although one iteration step requires more computational resources than the latter. 

We have tested the NFPT code based on the new method against the analytical FFF models by \citet{Titov99} with other available variational codes, our own VFVP code and the optimization code in SolarSoft (SSW). Our new NFPT code excels relative to others in reproducing characteristic features of TD models, for example, one flux rope with a proper twist, a bald patch with a proper writhe and quasi-separatrix layers with a hyperbolic flux tube. The NFPT code also produces the best scores in most performance metrics \citep{Schrijver06}, especially in $ CW{\mathrm {sin} } $ measuring the solution's own force-freeness. 
The application of the NFPT code has also been made to the vector magnetograms of a real active region, which will be reported in a sequel paper shortly.

\acknowledgments

The authors thank the reviewer for his/her constructive comments, which have helped improve the paper a lot. 
This work has been supported by the National Research Foundation of Korea (NRF) grants 2016R1D1A1B03936050 and 2019R1F1A1060887 funded by the Korean government.  
It was also partially supported by the Korea Astronomy and Space Science Institute under the R\&D program (Project No. 2017-1-850-07). 
Part of this work was performed when S.~Y.  and G.~S.~C. visited the Max-Planck-Institut f\"{u}r Sonnensystemforschung, G\"{o}ttingen, Germany, 
and they appreciate the fruitful interactions with researchers there. 

\vspace{5mm}
\software{ 
SolarSoft \citep{Freeland98}, 
FISHPACK \citep{Swarztrauber75}
          }

\bigbreak

\bigbreak

\bigbreak

\bigbreak

\bigbreak

\bigbreak

\appendix

\section{On the PT representation of a periodic field in Cartesian Coordinates}
\label{app:ptperiod}

Here we will plainly expound why a special treatment is needed in a PT representation 
in a Cartesian coordinate system when the magnetic field is periodic in two directions spanning a toroidal field surface. 
Our account here is a readable revision of previous studies \citep[e.g.,][and references therein]{Schmitt92}. 

Consider a magnetic field periodic in $x$ and $y$ with wavelengths (periods) $\lambda_x$ and $\lambda_y$, respectively. 
Taking $\xi=z$, we have 
\begin{align}
\label{eq:bx2}
& B_x = 
- { \partial \over {\partial x} } \left( { {\partial \Phi } \over {\partial z }  } \right) - { {\partial \Psi } \over {\partial y } } 
\, ,
\\
\label{eq:by2}
& B_y = 
- { \partial \over {\partial y} } \left( { {\partial \Phi } \over {\partial z }  } \right) + { {\partial \Psi } \over {\partial x } } 
\, .
\end{align}
We may define two vector fields ${\bm F}$ and ${\bm G}$ such that
\begin{align}
\label{eq:ffield}
{\bm F} & = F_x {\hat{\bm x}} +  F_y {\hat{\bm y}} 
= \Psi {\hat{\bm x}} + \left( - { {\partial \Phi } \over {\partial z }  } \right) {\hat{\bm y}}
\, ,
\\
\label{eq:gfield}
{\bm G} & = G_x {\hat{\bm x}} +  G_y {\hat{\bm y}} 
= \left( { {\partial \Phi } \over {\partial z }  } \right) {\hat{\bm x}} + \Psi {\hat{\bm y}} 
\, .
\end{align}
Note that ${\bm F}$ and ${\bm G}$ are also periodic in $x$ and $y$. 
We can then write 
\begin{align}
\label{eq:bx3}
B_x = {\hat{\bm z}} \cdot ( \nabla \times {\bm F} )
\, ,
\\
\label{eq:by3}
B_y = {\hat{\bm z}} \cdot ( \nabla \times {\bm G} )
\, .
\end{align}
We consider a 2D rectangular domain $ S_z = \{ (x, y, z) | 0 \le x \le \lambda_x, 0 \le y \le \lambda_y \} $ and apply Stokes' theorem
to have
\begin{align}
\label{eq:bxint}
\int_{S_z} B_x d{S_z} = \oint_{\partial S_z} {\bm F} \cdot d{\bm r}
\, ,
\\
\label{eq:byint}
\int_{S_z} B_y d{S_z}= \oint_{\partial S_z} {\bm G} \cdot d{\bm r}
\, .
\end{align}
In the line segments $x=0$ and $x=\lambda_x$, ${\bm F}(x,y,z)$ or ${\bm G}(x,y,z)$ is the same, but $d{\bm r}$ is in the opposite direction. 
The same is true for the line segments $y=0$ and $y=\lambda_x$. Thus, the contour integrals should be zero. 
However, the surface integrals may not be zero even if ${\bm B}$ is periodic. Such a field cannot
simply be expressed in the form of equations~(\ref{eq:bx2}) and (\ref{eq:by2}). In such cases, we define the following functions of $z$ only 
\begin{align}
\label{eq:bx0m}
B_{0x} (z) = {1 \over {\lambda_x \lambda_y} }  \int_{S_z} B_x \, d{S_z} 
\, ,
\\
\label{eq:by0m}
B_{0y} (z) = {1 \over {\lambda_x \lambda_y} }  \int_{S_z} B_y \, d{S_z} 
\, .
\end{align}
Then, the surface integrals of $B_x - B_{0x}$ and $B_y - B_{0y}$ are zero, and only ${\bm B} - B_{0x} {\hat{\bm x}} - B_{0x} {\hat{\bm x}}$ can 
be expressed by the standard PT representation in the form of equations~(\ref{eq:bx2}) and (\ref{eq:by2}). 

A similar argument applies to $B_z$ given by equation~(\ref{eq:bz}). Applying the divergence theorem to the 2D domain $S_z$, we have 
\begin{equation}
\label{eq:bzint}
\int_{S_z} B_z d{S_z} = \oint_{\partial S_z} { {\partial \Phi} \over {\partial n} } dl \, , 
\end{equation}
in which $dl = |d {\bm r}|$. 
For $\Phi$ periodic in $x$ and $y$, the contour integral is zero, but the surface integral of $B_z$ may not be zero even if $B_z$ is periodic. 
We then define 
\begin{equation}
\label{eq:bz0m}
B_{0z} =  {1 \over {\lambda_x \lambda_y} } \int_{S_z} B_z \, d{S_z} 
\, .
\end{equation}
Note that $B_{0z}$ is not a function of $z$, but a constant owing to the constraint $\nabla \cdot {\bm B} = 0$ 
combined with the periodicity of ${\bm B}$ in $x$ and $y$. 

To sum up, a magnetic field $\bm B$ periodic in Cartesian coordinates $x$ and $y$ is generally expressed as
\begin{equation}
{\bm B} =  \nabla \times ( {\hat{\bm z}} \times \nabla \Phi ) +  {\hat{\bm z}} \times \nabla \Psi  
+ {\bm B}_0
 \,  , 
\label{eq:special}
\end{equation}
in which 
\begin{equation}
{\bm B}_0
= B_{0x} (z) {\hat{\bm x}} + B_{0y} (z) {\hat{\bm y}} + B_{0z} {\hat{\bm z}} 
\label{eq:meanf}
\end{equation}
is so called a ``mean flow''  \citep{Schmitt92}. 
When periodic boundary conditions are explicitly imposed in two Cartesian coordinates, one should consider the mean flow. 
Since we do not employ a periodic boundary condition, a mean flow is not a consideration in our paper. 

\bigbreak 

 \section{On our earlier variational NLFFF code in vector potential formulation}
\label{app:var}

Before the NLFFF code based on a PT representation, we had developed and used a variational NLFFF code using a vector potential 
formulation of magnetic field. Since this code is based on a magnetofrictional method \citep{Chodura81}, the algorithm 
is as simple as 
\begin{equation}
\label{eq:dadt}
 { {\partial {\bm A}} \over {\partial t} } = -\nu ({\bm r}, t)  {\bm J}_\perp  \, ,
\end{equation}
in which $t$ is a pseudo-time, $\nu ({\bm r}, t)$ a proper coefficient maxizing the convergence rate, and 
$\displaystyle {\bm J}_\perp = {\bm B} \times ( {\bm J} \times {\bm B}  ) / {B^2}$. 
To expedite the convergence, we equip the code with a gradient descent algorithm \citep{Chodura81}.
When a vector potential  ${\bm A}$ is used to describe the magnetic field, we cannot set all three components of ${\bm A}$ at $z=0$ fixed 
in order to 
impose $B_z$ and $J_z$ there. 
In a magnetofrictional code by \citet{Roumeliotis96}, $A_z (x,y,0)$ was set fixed and $A_x (x,y,0)$ and $A_y (x,y,0)$ were varied at every time-step. 
In our variational code, we set 
$A_x (x,y,0)$ and $A_y (x,y,0)$ once for all to fix $B_z (x,y,0)$ and the solution of the following 2D Poisson equation is given 
as $A_z (x,y,0)$ at every time-step,  
\begin{equation}
\label{eq:azpoisson}
\nabla_{xy}^2 A_z |_{z=0} = - J_{z, {\rm obs}} + \left[ 
{\partial \over {\partial z}} \nabla \cdot {\bm A}_{xy}     \right]_{z=0^+} \, ,
\end{equation}
in which $J_{z, {\rm obs}} $ is the boundary condition of $J_z$ derived from an observation and the $z$-derivative is evaluated using a one-side finite differencing.
The computed $J_z(x,y,0)$ is thus equated 
with  $J_{z, {\rm obs}} $ at every time-step. 
Our variational code is working very well for moderately sheared fields \citep[e.g.,][]{Low90}, 
but shows a little weakness for magnetic fields with flux ropes as
with other variational codes. 
This motivated us to devise the new formulation presented in this paper, in which the imposition of the bottom boundary condition is tidy and effective.

\bigskip
\nolinenumbers
%
%

\end{document}